\documentclass[numberedappendix]{emulateapj}
\usepackage{apjfonts}
\slugcomment{Accepted by ApJ, \today}
\shorttitle{Deficit of wide binaries in $\eta$\,Cha}
\shortauthors{Brandeker et al.}

\begin{document}

\title{Deficit of wide binaries in the $\eta$~Chamaeleontis young cluster}

\author{Alexis Brandeker\altaffilmark{1}, Ray Jayawardhana, Parandis Khavari}
\affil{Department of Astronomy and Astrophysics, University of Toronto,
    50 St.~George Street, Toronto, ON M5S~3H4, Canada}
\email{brandeker@astro.utoronto.ca}

\author{Karl E.\ Haisch, Jr.\ }
\affil{Physics Department, Utah Valley State College, 
800 West University Parkway, Orem, UT 84058-5999}

\and

\author{Diego Mardones}
\affil{Departamento de Astronom\'ia, Universidad de Chile, Casilla 36-D, Santiago, Chile}
\altaffiltext{1}{Stockholm Observatory, AlbaNova University Centre, SE-106~91 Stockholm, Sweden}

\newtheorem{theorem}{Theorem}{\bf}{\it}
\newtheorem{proposition}[theorem]{Proposition}{\bf}{\it}
\newtheorem{lemma}[theorem]{Lemma}{\bf}{\it}

\begin{abstract}
We have carried out a sensitive high-resolution imaging survey of stars in the
young (6--8\,Myr), nearby (97\,pc) compact cluster around $\eta$~Chamaeleontis to
search for stellar and sub-stellar companions. Given its youth and proximity,
any sub-stellar companions are expected to be luminous, especially in the near
infrared, and thus easier to detect next to their parent stars. Here, we present
VLT/NACO adaptive optics imaging with companion detection limits for 17 $\eta$\,Cha
cluster members, and follow-up VLT/ISAAC near-infrared spectroscopy for companion
candidates. The widest binary detected is $\sim$0\farcs2, corresponding to the projected
separation 20\,AU, despite our survey being sensitive down to sub-stellar
companions outside 0\farcs3, and planetary mass objects outside 0\farcs5. This implies
that the stellar companion probability outside 0\farcs3 and the brown dwarf companion
probability outside 0\farcs5 are less than 0.16 with 95\,\% confidence. We compare the
wide binary frequency of $\eta$\,Cha to that of the similarly aged TW~Hydrae
association, and estimate the statistical likelihood that the wide binary
probability is equal in both groups to be $< 2 \times 10^{-4}$. Even though the 
$\eta$\,Cha cluster is relatively dense, stellar encounters in its present
configuration cannot account for the relative deficit of wide binaries. We thus conclude
that the difference in wide binary probability in these two groups provides strong evidence
for multiplicity properties being dependent on environment. In two appendices we
derive the projected separation probability distribution for binaries, used to
constrain physical separations from observed projected separations, and summarize
statistical tools useful for multiplicity studies.
\end{abstract}




\section{Introduction}

Most solar-type stars reside in binaries \citep{duq91}, yet their formation is 
not well understood. It is generally accepted
that most stars form in groups rather than in isolation \citep{ada01}. Intriguingly,
the multiplicity of young stars in many regions seems to be systematically higher 
than that of their 
main-sequence counterparts \citep[e.g.][and references therein]{mat00}. The
reason for this multiplicity overabundance is not entirely understood, although
two principal scenarios have been proposed. 
One is that the multiplicity fraction is sensitive to the initial conditions of
the star formation process. That would imply that different regions show a 
variety of multiplicity properties, and indeed, there seems to be some evidence 
for the multiplicity fraction being anticorrelated with the stellar density of
the region \citep{pat01}.
Another possibility is that young stars start out with a high fraction of multiples
that subsequently are disrupted due to dynamical evolution \citep[e.g.][]{rei00}. 
Star-forming regions
are generally too dispersed for binaries to be disrupted by interactions between
members. The alternative is that many stars form not just in binaries, but in 
unstable higher-order multiples that get disrupted with time, as indeed suggested by 
numerical simulations \citep{bat02,del04}. For a recent review on the current 
status of the field, see \citet{duc06}.

The nearby (97\,pc), young \citep[6--8\,Myr,][and references therein]{jil05} 
$\eta$\,Chamaeleontis cluster was 
found by the X-rays emitted from its members, as revealed by \textit{ROSAT}, 
together with their common space motion, as revealed by \textit{Hipparcos} \citep{mam99}.
The cluster is mainly populated by late-type stars (K3-M6, see Table~\ref{t:targets}), 
and shows no evidence for extinction \citep{mam00}. These properties all contribute
to make the cluster an excellent laboratory for investigating brown dwarf (BD) and
planet formation, and their evolution \citep[e.g.][]{son04,luh04a,hai05,lyo06,jay06}.
In particular, using adaptive optics (AO) systems on large telescopes, it is 
possible to reach high contrast-ratio sensitivities close in, to detect any 
wide ($>$0\farcs5) companion down to planetary masses.

In a previous high-resolution survey using speckle interferometry and AO
on 2.2--3.6\,m telescopes, \citet{koh02} searched 13 members for companions
($\eta$\,Cha~1--12 and 15, see Table~\ref{t:targets}). They found two resolved 
binaries, $\eta$\,Cha~1 and $\eta$\,Cha~9, 
and one suspected unresolved binary, $\eta$\,Cha~12.
$\eta$\,Cha~12 has also been suspected to be binary due to its
elevation in color-magnitude diagrams \citep{law01,luh04b}.

In the present paper we report on a deep AO search for faint companions to 17 
confirmed members of the $\eta$\,Cha cluster. \S\,\ref{s:obs}
details the observations. \S\,\ref{s:analysis} describes analysis and
results, including how astrometry and photometry were performed, how
the Strehl ratio was measured and the contrast sensitivity estimated.
In \S\,\ref{s:discuss} we discuss the companion candidates and constrain 
the orbit of the previously observed 0\farcs2 
binary $\eta$\,Cha~1 (Fig.~\ref{f:ec01}).
We then go on to determine limits on the companion probability, use
this to discuss the apparent deficit of binaries in the $\eta$\,Cha
cluster, and estimate the likelihood of binaries being disrupted by
stellar encounters. \S\,\ref{s:conclusions} contains an enumerated list 
of conclusions.

In the Appendices, we first derive the (time-averaged) probability distribution 
for the projected separation of a binary, given its semi-major axis. We
then continue with some useful properties of multiplicity statistics, in particular
how to estimate binomial confidence intervals, and how to test if two outcomes
are derived from the same binomial distribution.


\section{Observations}
\label{s:obs}
%
\begin{deluxetable*}{clcccccl}
\tabletypesize{\scriptsize}
\tablewidth{0pt}
\tablecaption{\label{t:targets}Targets observed in the $\eta$~Chamaeleontis cluster}
\tablehead{
\colhead{$\eta$\,Cha\tablenotemark{a}} & 
\colhead{Name} & 
\colhead{$\alpha_{\mathrm{J}2000.0}$} & 
\colhead{$\delta_{\mathrm{J}2000.0}$} & 
\colhead{$J$} & 
\colhead{$H$} & 
\colhead{$K_s$} & 
\colhead{SpT}
}
\startdata
\phn1 & EG\,Cha &         08:36:56.24 & -78:56:45.5 & \phn8.155\,$\pm$\,0.019 & \phn7.498\,$\pm$\,0.049 & \phn7.338\,$\pm$\,0.021 & K4   \\
\phn2 & $\eta$\,Cha &     08:41:19.48 & -78:57:48.1 & \phn5.688\,$\pm$\,0.019 & \phn5.721\,$\pm$\,0.040 & \phn5.718\,$\pm$\,0.018 & B8   \\
\phn3 & EH\,Cha &         08:41:37.03 & -79:03:30.4 &    10.349\,$\pm$\,0.023 & \phn9.647\,$\pm$\,0.022 & \phn9.415\,$\pm$\,0.019 & M3   \\
\phn4 & EI\,Cha &         08:42:23.72 & -79:04:03.0 & \phn9.535\,$\pm$\,0.024 & \phn8.779\,$\pm$\,0.061 & \phn8.615\,$\pm$\,0.019 & K7   \\
\phn5 & EK\,Cha &         08:42:27.11 & -78:57:47.9 &    10.777\,$\pm$\,0.023 &    10.099\,$\pm$\,0.021 & \phn9.855\,$\pm$\,0.021 & M5   \\
\phn6 & EL\,Cha &         08:42:38.80 & -78:54:42.8 &    10.232\,$\pm$\,0.027 & \phn9.584\,$\pm$\,0.023 & \phn9.290\,$\pm$\,0.021 & M2   \\
\phn7 & EM\,Cha &         08:43:07.24 & -79:04:52.5 & \phn8.420\,$\pm$\,0.024 & \phn7.758\,$\pm$\,0.034 & \phn7.635\,$\pm$\,0.033 & K3   \\
\phn8 & RS\,Cha &         08:43:12.23 & -79:04:12.3 & \phn5.994\,$\pm$\,0.030 & \phn5.877\,$\pm$\,0.038 & \phn5.852\,$\pm$\,0.034 & A7   \\
\phn9 & EN\,Cha &         08:44:16.38 & -78:59:08.1 &    10.260\,$\pm$\,0.026 & \phn9.668\,$\pm$\,0.026 & \phn9.335\,$\pm$\,0.024 & M4   \\
10 &    EO\,Cha &         08:44:31.88 & -78:46:31.2 & \phn9.653\,$\pm$\,0.023 & \phn8.919\,$\pm$\,0.063 & \phn8.732\,$\pm$\,0.021 & K7   \\
11 &    EP\,Cha &         08:47:01.66 & -78:59:34.5 & \phn8.729\,$\pm$\,0.020 & \phn8.025\,$\pm$\,0.055 & \phn7.655\,$\pm$\,0.038 & K4   \\
12 &    EQ\,Cha &         08:47:56.77 & -78:54:53.2 & \phn9.323\,$\pm$\,0.024 & \phn8.683\,$\pm$\,0.082 & \phn8.410\,$\pm$\,0.031 & M1   \\
13 &    HD\,75505       & 08:41:44.72 & -79:02:53.1 & \phn7.059\,$\pm$\,0.026 & \phn6.987\,$\pm$\,0.036 & \phn6.928\,$\pm$\,0.023 & A1V  \\
15 & ECHA\,J0843.3-7905 & 08:43:18.58 & -79:05:18.2 &    10.505\,$\pm$\,0.026 & \phn9.834\,$\pm$\,0.021 & \phn9.431\,$\pm$\,0.023 & M4   \\
16 & ECHA\,J0844.2-7833 & 08:44:09.15 & -78:33:45.7 &    12.505\,$\pm$\,0.024 &    11.976\,$\pm$\,0.022 &    11.618\,$\pm$\,0.024 & M5   \\
17 & ECHA\,J0838.9-7916 & 08:38:51.50 & -79:16:13.7 &    11.275\,$\pm$\,0.023 &    10.721\,$\pm$\,0.022 &    10.428\,$\pm$\,0.023 & M5   \\
18 & ECHA\,J0836.2-7908 & 08:36:10.73 & -79:08:18.4 &    11.849\,$\pm$\,0.024 &    11.277\,$\pm$\,0.026 &    10.945\,$\pm$\,0.021 & M5.5 
\enddata
\tablecomments{Coordinates and IR magnitudes are from the 2MASS All Sky Data Release.
The spectral types are from \citet{mam99} for 1--12, \citet{hou75} for 13, \citet{law02} for 15, 
and \citet{son04} for 16--18.}
\tablenotetext{a}{Numbers 1--12 coincide with the RECX numbers introduced by \cite{mam99}, 
and 1--18 with \cite{luh04b}. Their number 14 (``USNO Anon 1'') was not observed.}
\end{deluxetable*}
\begin{deluxetable}{lclrr}
\tablewidth{0pt}
\tabletypesize{\scriptsize}
\tablecaption{\label{t:obslog}Observation log}
\tablehead{
\colhead{} & 
\colhead{} & 
\colhead{} & 
\colhead{DIT} & 
\colhead{Total} \\
\colhead{$\eta$\,Cha\tablenotemark{a}} & 
\colhead{UT Date} & 
\colhead{Instrument\tablenotemark{b}} & 
\colhead{(s)} & 
\colhead{(min)}
}
\startdata
\phn1     & 2004-04-03 & VIS $H$ cor.    &   3    & 19   \\
\phn2     & 2004-04-04 & VIS $H$ cor.    &   3    & 15   \\
\phn3     & 2004-04-02 & N90C10 $J$      &  20    & 13   \\
\phn3     & 2002-10-20 & VIS $H$         &   0.6  & 15   \\
\phn3     & 2002-10-21 & VIS $H$         &   0.6  & 14   \\
\phn3     & 2004-04-02 & N90C10 $K_s$    &  20    & 27   \\
\phn4\,B  & 2004-04-22 & ISAAC $H$ spec. & 200    & 40   \\
\phn4     & 2002-11-17 & VIS $H$ ND      &   1.5  &  0.7 \\
\phn4     & 2002-11-17 & VIS $H$         &   0.34 & 13   \\
\phn5     & 2003-01-17 & N20C80 $H$      &   0.5  & 20   \\
\phn5     & 2003-01-21 & N20C80 $H$      &   0.5  & 14   \\
\phn5     & 2003-02-17 & N20C80 $H$      &   0.5  &  6   \\
\phn5     & 2003-02-19 & N20C80 $H$      &   0.5  & 13   \\
\phn6     & 2003-01-18 & VIS $H$ ND      &   5    &  1.6 \\
\phn6     & 2003-01-18 & VIS $H$         &   0.35 & 21   \\
\phn7     & 2004-04-04 & VIS $H$ cor.    &  20    & 20   \\
\phn8     & 2004-04-04 & VIS $H$ cor.    &   3    & 14   \\
\phn9     & 2004-04-02 & N90C10 $J$      &   8    & 13   \\
\phn9     & 2003-01-21 & VIS $H$ ND      &   6    &  2   \\
\phn9     & 2003-01-21 & VIS $H$         &   0.35 & 13   \\
\phn9     & 2004-04-02 & N90C10 $K_s$    &  16    & 27   \\
\phn9\,Aab  & 2004-04-03 & N90C10 $H$ spec.& 120    & 40 \\
\phn9\,B  & 2004-04-22 & ISAAC $H$ spec. & 200    & 47   \\
   10     & 2003-01-22 & VIS $H$ ND      &   1.5  &  0.5 \\
   10     & 2003-01-22 & VIS $H$         &   0.35 & 13   \\
   10\,B  & 2004-04-24 & ISAAC $H$ spec. & 200    & 60   \\
   11     & 2003-01-22 & VIS $H$ ND      &   1.2  &  0.4 \\
   11     & 2003-01-22 & VIS $H$         &   0.35 & 13   \\
   12     & 2003-01-22 & VIS $H$ ND      &   1.5  &  0.5 \\
   12     & 2003-01-22 & VIS $H$         &   0.35 & 13   \\
   13     & 2004-04-03 & VIS $H$ cor.    &  10    & 12   \\
   15     & 2004-04-02 & N90C10 $J$      &  16    & 27   \\
   15     & 2003-01-21 & VIS $H$ ND      &   6    &  4   \\
   15     & 2003-01-21 & VIS $H$         &   0.35 & 11   \\
   15     & 2003-01-22 & VIS $H$         &   0.35 & 11   \\
   15     & 2004-04-02 & N90C10 $K_s$    &  16    & 26   \\
   15\,C  & 2004-04-22 & ISAAC $H$ spec. & 200    & 40   \\
   16     & 2004-04-02 & N90C10 $H$      &  80    & 36   \\
   17     & 2004-04-03 & N90C10 $H$      &  10    & 33   \\
   18     & 2004-04-02 & N90C10 $H$      &  30    & 33   
\enddata
\tablenotetext{a}{\ Letters refer to the companion candidate
 observed in addition to the primary.}
\tablenotetext{b}{For NACO observations, VIS, N20C80 and N90C10 correspond
 to the used WFS, $J$, $H$, and $K_s$ are the used filters, ND means neutral density
filter, cor.\ coronographic observations, and spec.\ spectroscopy. All ISAAC
observation were made in the same mode (see \S\,\ref{s:obs}).}
\end{deluxetable}
\begin{figure}
\includegraphics[width=\hsize,clip=true]{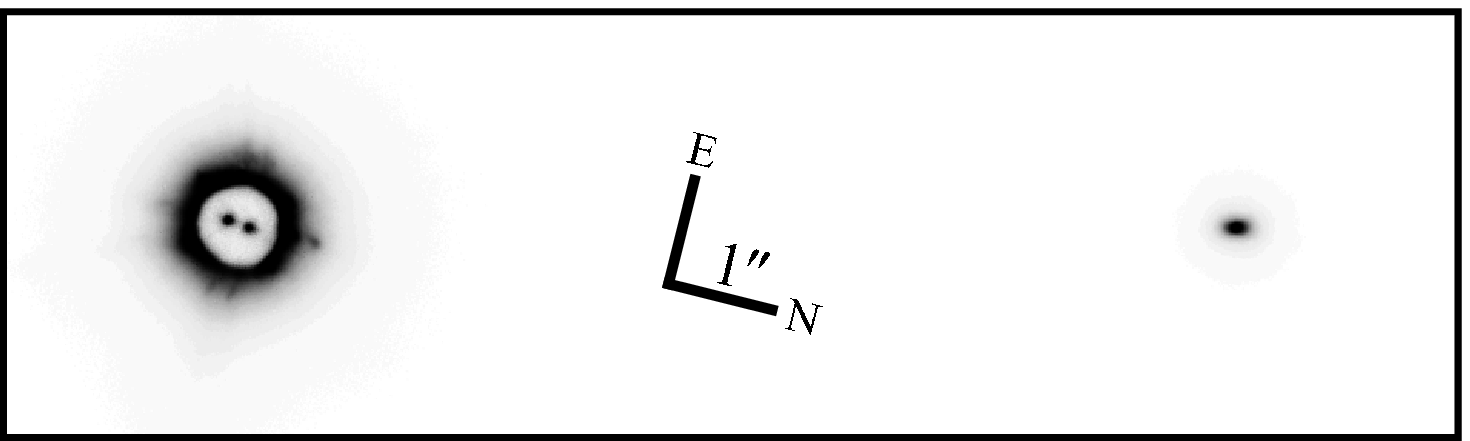}
\caption{\label{f:ec01}The close 0\farcs2 binary $\eta$\,Cha~1 observed with a
semi-transparent coronographic mask, and a likely background object 9\arcsec\ away
(\S\,\ref{s:companions}).
The orientation and scale are shown by the two 1\arcsec\ perpendicular axes..}
\end{figure}
\begin{figure}
\includegraphics[width=\hsize,clip=true]{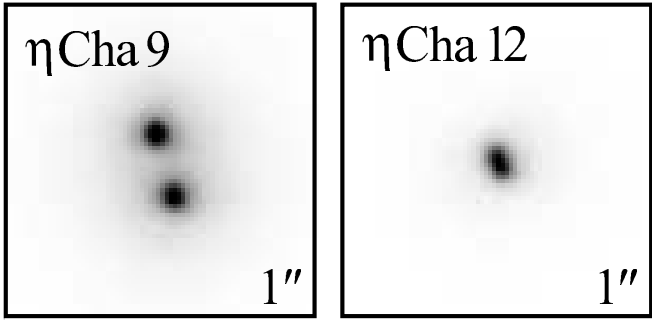}
\caption{\label{f:ec0912}Close-ups of the tight binaries $\eta$\,Cha~9 (left) and
$\eta$\,Cha~12 (right). The box sizes are 1\arcsec\ on a side. $\eta$\,Cha~9 is a well
resolved 0\farcs2 binary, while $\eta$\,Cha~12 is unresolved. From the asymmetric
profile, we infer that the $\eta$\,Cha~12 separation is $\sim$0\farcs04 
(\S\,\ref{s:astrophot}). North is up, and east is to the left.}
\end{figure}
We have observed 17 confirmed members of the $\eta$\,Cha cluster with 
high-resolution AO imaging and follow-up spectroscopy of companion candidates.
Table~\ref{t:targets} summarizes the observed target parameters and assigns a
running target identification number used in this paper.
Imaging data were obtained in service mode with the AO system NACO 
on the 8.2\,m VLT Yepun at ESO (Cerro Paranal, Chile), during two different 
semesters (2002/2003 and 2004). Spectroscopic follow-up observations were obtained
in service mode during 2004 using NACO and ISAAC (on VLT Antu).

For imaging with NACO, we used the high-resolution lens, giving
a pixel scale of 13.26\,mas\,pix$^{-1}$\ and the field of view 
13\farcs6$\times$13\farcs6. The full-width half maximums (FWHMs) of the diffraction 
limited point-spread functions (PSFs) of the setup in our used 
photometric bands $J$ (1.27\,$\mu$m), $H$ (1.66\,$\mu$m), and $K_s$ (2.18\,$\mu$m)
are 32\,mas, 42\,mas, and 56\,mas, respectively.
In the first semester, we obtained an extra exposure with a 5~mag neutral 
density filter for the stars we expected would saturate the array during 
the minimum exposure time of 0.3454\,s.
In the second semester, we instead made use of the new semi-transparent 
coronographic mask for the brightest targets, that reduces the light
within 0\farcs35 radius. We measured the contrast between the inside and outside
of the mask to be $6.0\pm0.1$~mag in $H$ and  $6.3\pm0.1$~mag in $K_s$, by observing 
a binary with and without the coronographic plate. This is consistent with results 
found by the NACO instrument team (N.~Ageorges, private communication). 
We used either the visual wave front sensor (WFS) mode VIS,
or the infrared WFS modes N20C80 or N90C10, that direct 20\%
or 90\% of the IR light to the WFS, and the rest to the
science camera CONICA.
The zenith seeing in the $V$-band was better than 0\farcs6 in general, though the 
targets, due to their low declination of -78\degr, were observed at the relatively 
high air mass of 1.5--2.0.
Three examples of obtained images are show in 
Figs.~\ref{f:ec01}\,\&\,\ref{f:ec0912}.

For spectroscopy with NACO, we used Grism~3 with the S27 lens, the N90C10
WFS, and a 0\farcs086$\times$40\arcsec\ slit that produces an $H$-band spectrum
from 1.44\,$\mu$m to 1.72\,$\mu$m at spectral resolution $R\sim1500$
and pixel scale 27\,mas\,pix$^{-1} \times$ 3.4\,\AA\,pix$^{-1}$.

For spectroscopy with ISAAC, we used the SWS1-LR mode with the $SH$ filter,
and a 1\arcsec$\times$120\arcsec\ slit that produces an $H$-band spectrum
from 1.48\,$\mu$m to 1.80\,$\mu$m at spectral resolution $R\sim500$
and pixel scale 0\farcs148\,mas\,pix$^{-1} \times$ 4.1\,\AA\,pix$^{-1}$.

The basic reduction was done in a standard way, making use of the 
reduction pipeline in the case of ISAAC. The sky was estimated
from the jittered observations and subtracted from all frames, which were 
subsequently corrected for cosmic rays and flat fielded. Since the coronographic
mask is in a fixed position on the array, half of the integration time was spent 
with the source chopped out of the field. For the spectroscopic observations,
the slit was put over both primary and companion candidate, and then jittered 
along the slit. To decompose the spectra of two stars on the slit, we extracted 
the spectra by fitting two-component Moffat functions \citep{mof69} to the 
spatial profiles. The wavelength calibration made use of Ar lamp spectra obtained
during daytime. To correct for atmospheric absorption lines, an early type
(B2--B5) telluric standard star was observed each night, at a similar airmass (within 0.2) 
to the target. The science spectra were then divided by the telluric standard spectra, 
multiplied by standard star models, and normalized. Unfortunately, the telluric line
correction proved to not be entirely reliable, most probably due to the high airmass
($\sim$2) at which the observations were made.

The use of AO and a narrow slit makes the calibration of NACO spectra 
difficult. For example, the PSF, and therefore slit loss, varies 
with wavelength. In the NACO $H$-band spectra of the resolved $\eta$\,Cha~9 binary we 
noticed the spectral shape to be somewhat steeper than the ISAAC $H$-band spectrum 
we have for the unresolved binary. Since the observations were made at high airmass, 
we suspect this 
difference might be due to additional wavelength-dependent slit losses, caused by 
atmospheric differential refraction. To test this hypothesis we computed the
expected atmospheric dispersion using the refraction index from \citet{pec72}, 
and the standard dispersion equation (e.g., equation~3 in \citealt{roe02}),
giving the expected dispersion of 50\,mas between 1.5\,$\mu$m and 1.7\,$\mu$m,
which is a fair fraction of the 86\,mas slitwidth. Projected on the slit orientation,
the computed offset increases from 25\,mas to 34\,mas during the 10 exposures of
$\eta$\,Cha~9, while at the same time the observed spectral slope gets steeper by
$\sim$15\%, indicating that differential refraction may indeed be significant. The
telluric standard was observed with the slit at the parallactic angle, and was thus not
affected by differential refraction.
These wavelength-dependent slit losses are of no consequence for the present observations, 
since we are interested in the spectral difference between the two components, but they
might be of importance for future observations using VLT/NACO at high airmass.

Table~\ref{t:obslog} presents the observing log, with epochs, instrumentation,
exposure time, and total integration time on source for each target.


\section{Analysis and results}
\label{s:analysis}
%
\subsection{Astrometry and photometry of sources in the field}
\label{s:astrophot}
%
\begin{deluxetable}{ccccc}
\tablewidth{0pt}
\tablecaption{\label{t:comp}Confirmed companions}
\tablehead{
\colhead{$\eta$\,Cha} & 
\colhead{Sep (\arcsec)} & 
\colhead{PA\tablenotemark{a} (\degr)} & 
\colhead{Band} &
\colhead{FR\tablenotemark{b}}
}
\startdata
1 & 0.192\,$\pm$\,0.002 &    354.0\,$\pm$\,0.3 & $H$   & \phn1.15\,$\pm$\,0.02 \\
9 & 0.204\,$\pm$\,0.003 &    195.7\,$\pm$\,0.5 & $J$   & \phn1.03\,$\pm$\,0.04 \\
9 & 0.207\,$\pm$\,0.003 &    195.8\,$\pm$\,0.5 & $H$   & \phn0.96\,$\pm$\,0.04 \\
9 & 0.209\,$\pm$\,0.003 &    196.4\,$\pm$\,0.5 & $K_s$ & \phn1.05\,$\pm$\,0.04 \\
12\tablenotemark{c} & 0.040\,$\pm$\,0.010 & 28\,$\pm$\,4   & $H$   & \phn\phn1.0\,$\pm$\,0.1
\enddata
\tablenotetext{a}{The position angle is measured from north to east. There is an additional systematic error of $\sim1\degr$.}
\tablenotetext{b}{The flux ratio A/B.}
\tablenotetext{c}{$\eta$\,Cha\,12 is unresolved, see \S\,\ref{s:astrophot}.}

\end{deluxetable}

\begin{deluxetable}{ccccc}
\tablewidth{0pt}
\tablecaption{\label{t:bg}Likely chance alignment objects}
\tablehead{
\colhead{$\eta$\,Cha} & 
\colhead{Sep (\arcsec)} & 
\colhead{PA\tablenotemark{a} (\degr)} & 
\colhead{Band} &
\colhead{Mag\tablenotemark{b}}
}
\startdata
  \phn\phm{$^c$}1\tablenotemark{c} & 8.610\,$\pm$\,0.018 &  \phn13.8\,$\pm$\,0.2 & $H$\phn & 11.5\,$\pm$\,1.0 \\
  \phn2 & 7.997\,$\pm$\,0.020 &     263.6\,$\pm$\,0.2 & $H$\phn & 19.8\,$\pm$\,0.5 \\
  \phn3 & 2.084\,$\pm$\,0.005 &     106.5\,$\pm$\,0.3 & $J$\phn & 17.0\,$\pm$\,0.5 \\
  \phn3 & 2.034\,$\pm$\,0.005 &     106.6\,$\pm$\,0.3 & $H$\phn & 16.3\,$\pm$\,0.2 \\
  \phn3 & 2.082\,$\pm$\,0.005 &     106.1\,$\pm$\,0.3 & $K_s$   & 15.9\,$\pm$\,0.2 \\
  \phn\phm{$^i$}4\tablenotemark{d} & 7.352\,$\pm$\,0.018 &     273.3\,$\pm$\,0.2 & $H$\phn & 13.9\,$\pm$\,0.3 \\
  \phn7 & 5.611\,$\pm$\,0.016 &     344.2\,$\pm$\,0.2 & $H$\phn & 15.9\,$\pm$\,1.0 \\
  \phn8 & 9.443\,$\pm$\,0.023 &  \phn30.8\,$\pm$\,0.2 & $H$\phn & 17.4\,$\pm$\,1.0 \\
 \phn\phm{$^e$}9\tablenotemark{e} & 3.582\,$\pm$\,0.010 &     141.7\,$\pm$\,0.3 & $J$\phn & 15.6\,$\pm$\,0.2 \\
 \phn\phm{$^e$}9\tablenotemark{e} & 3.526\,$\pm$\,0.010 &     142.2\,$\pm$\,0.3 & $H$\phn & 15.3\,$\pm$\,0.1 \\
 \phn\phm{$^e$}9\tablenotemark{e} & 3.568\,$\pm$\,0.010 &     141.8\,$\pm$\,0.3 & $K_s$   & 15.1\,$\pm$\,0.1 \\
 10 & 9.895\,$\pm$\,0.026 &  \phn62.5\,$\pm$\,0.2 & $H$\phn & 16.6\,$\pm$\,0.5 \\
  \phm{$^i$}15\tablenotemark{f} & 2.707\,$\pm$\,0.040 &  \phn73.4\,$\pm$\,0.8 & $J$\phn & 18.0\,$\pm$\,0.5 \\
 15 & 2.726\,$\pm$\,0.008 &  \phn72.1\,$\pm$\,0.3 & $H$\phn & 17.5\,$\pm$\,0.2 \\
 \phm{$^i$}15\tablenotemark{f} & 2.752\,$\pm$\,0.040 &  \phn72.7\,$\pm$\,0.8  & $K_s$   & 17.3\,$\pm$\,0.5 \\
 \phm{$^g$}15\tablenotemark{g} & 6.370\,$\pm$\,0.017 &     209.4\,$\pm$\,0.2 & $J$\phn & 13.8\,$\pm$\,0.4 \\
 15 & 6.352\,$\pm$\,0.017 &     210.0\,$\pm$\,0.2 & $H$\phn & 13.6\,$\pm$\,0.2
\enddata
\tablenotetext{a}{The position angle is measured from north to east. There is an additional systematic error of $\sim1\degr$.}
\tablenotetext{b}{Apparent magnitudes were derived from Table~\ref{t:targets}.}
\tablenotetext{c}{The astrometry was measured relative to the photocenter of the inner binary. This object
 is listed in 2MASS with $J=11.70\pm0.06$, $H=11.12\pm0.08$, and $K=11.06\pm0.06$, and in DENIS with
 $I = 12.43\pm0.03$.}
\tablenotetext{d}{This object
 is listed in 2MASS as having $J=14.48\pm0.23$, $H=13.81\pm0.32$, and $K=13.72\pm0.12$.}
\tablenotetext{e}{The astrometry was measured relative to the south component B of the inner binary.}
\tablenotetext{f}{Marginal detection}
\tablenotetext{g}{Object located at edge of array.}

\end{deluxetable}
For observations where multiple sources were well separated in the field, we
measured the pixel positions on the array by making use of the \texttt{iraf} 
routine \texttt{imexamine}. To determine the relative precision of the astrometry
we measured the separation between the tight, but well resolved, binary $\eta$\,Cha~1
for 10 different consecutive frames. The standard deviation was found to be about
0.12 pixels, or $\sigma_{\mathrm{std}} = 1.6$\,mas. In addition, there is a systematic 
image scale uncertainty of $\sigma_{\mathrm{scl}} \sim 0.01$\,mas\,pix$^{-1}$, so the 
estimated total separation error $\sigma_{\mathrm{sep}}$ depends on the separation $s$ as 
$\sigma_{\mathrm{sep}}(s) = [(\sigma_{\mathrm{scl}}s)^2 + \sigma_{\mathrm{std}}^2]^{1/2}$. For
some of the fainter sources, $\sigma_{\mathrm{std}}$ would be larger by a factor of several,
due to poor centering. 

The relative position angle error $\sigma_{\mathrm{PA}}$ was computed from 
$\sigma_{\mathrm{std}}$ by scaling the error with separation, i.e.\ $\sigma_{\mathrm{PA}}(s) = 
\arctan(\sigma_{\mathrm{std}} s^{-1}) \approx \sigma_{\mathrm{std}} s^{-1}$\,rad.
The position angle error is generally dominated by the systematic error of $\sim 1\degr$, due to the 
uncertain orientation of the array. The astrometry is summarized in Table~\ref{t:comp}\,\&\,\ref{t:bg}.

Photometric measurements of AO data are complicated by the spatially 
varying PSF. To estimate the level of anisoplanatism, we measured the PSF
for the targets with multiple sources in the field. The FWHM
of the PSF core (and thus the Strehl ratio) was found to be strongly dependent 
on the distance to the primary WFS. Out to a few arcseconds, the
variation is less than 10\%, but at larger separations ($\ga$7\arcsec) 
the Strehl ratio can decrease by as much as a factor of 10. Part of the reason for this
strong anisoplanatism may be that the stars were all observed at relatively high
airmass ($\eta$\,Cha never rises below an airmass of 1.5 from Cerro Paranal).

Because of the dependence of the PSF on separation, sources separated
$\ga$4\arcsec\ from the primary have poorly constrained photometry. For those
well-separated sources, we used a large aperture of 100 pixel radius for both the 
primary and secondary, if it was bright enough; otherwise
we fit a Gaussian to the PSF core and computed the integrated flux under the
Gaussian. Since the Strehl ratio is so low for the sources at large separation from
the WFS star, their PSFs are reasonably well approximated by Gaussians. By using the
Gauss-fitting procedure also on the bright wide companions, we found the two
methods to be consistent within 0.3\,mag, which is the quoted error. As a second 
consistency check we used the relatively 
bright source 7\arcsec\ from $\eta$\,Cha~4, catalogued as $H = 13.81 \pm 0.32$
by 2MASS, in agreement with our estimated $H = 13.9 \pm 0.3$.

For the 
close-in companions, where the PSF is roughly constant, we used aperture
photometry with a radius of 4 pixels, and subtracted the background halo from 
the primary by measuring its brightness at the same separation but opposite 
position angle. The photometry is much more robust in this case. The error
of the measured flux ratios is estimated by finding the consistency of
multiple (consecutive) observations of the same target.

As the binary $\eta$\,Cha~12 is unresolved (Fig.~\ref{f:ec0912}), it requires 
special attention.
We assume that the PSF core of the observation is circularly symmetric, 
as is the case for the other observed central PSFs, and that the elongation to the north
east is the result of an equal mass binary. The position angle is then estimated
from the position angle of an elliptical Gaussian fit, and the separation from
the FWHM of the fit along and across the major axis, by computing their 
difference.

\subsection{Strehl ratios\label{s:strehl}}
The Strehl ratio $S$ is defined as the ratio between the observed peak intensity
$P_{\mathrm{obs}}$ of the PSF, and the theoretical peak intensity 
$P_{\mathrm{ideal}}$ of an idealized telescope (no distortion of wavefront, 
no obstructions) of the same aperture observing the same star, 
$S \equiv P_{\mathrm{obs}} / P_{\mathrm{ideal}}$. Instead of computing
$P_{\mathrm{ideal}}$ for each observation, we estimated the ratio 
$R_{\mathrm{ideal}} = P_{\mathrm{ideal}} / A_{\mathrm{ideal}}$, where 
$A_{\mathrm{ideal}}$ is the flux (integrated intensity), that only depends 
on telescope aperture and wavelength. For an 8.2\,m telescope we computed
$R_J$ = 772\,arcsec$^{-2}$, $R_H$ = 452\,arcsec$^{-2}$, and $R_{K_s}$ = 263\,arcsec$^{-2}$, 
for $J$, $H$, and $K_s$, respectively.
We then estimated similar ratios
$R_{\mathrm{obs}} = P_{\mathrm{obs}} / A_{\mathrm{obs}}$ for the observations, and
determined the Strehl ratio as $S =  R_{\mathrm{obs}} / R_{\mathrm{ideal}}$. The
peak intensity of the observed profile was found
by fitting a circular Gaussian function to the PSF core, and the integrated 
intensity by summing up all pixels within a circular aperture of 2\arcsec\ radius.

The estimated Strehl ratios are shown in Figs.~\ref{f:contrast1}\,\&\,\ref{f:contrast2},
except for the coronographic observations where we could not determine $R_{\mathrm{obs}}$ reliably,
due to the presence of the mask. Because the coronographic targets are bright, and observed
under similar conditions, we expect the Strehl ratios to be at the high end ($>$15\,\%). 

The lower the Strehl ratio, the larger fraction of the stellar flux that is diluted into the
seeing disc. Even down to Strehl ratios of a few percent, however, there usually is a
diffraction-limited core of the PSF. This means that searches for point sources (such as 
stars) are greatly aided by AO even at low Strehl ratios, while searches for extended 
structures (such as circumstellar material) are critically dependent on high Strehl 
ratios.

\subsection{Contrast sensitivities}
\label{s:contrast}
\begin{figure*}
\includegraphics[width=\hsize,clip=true]{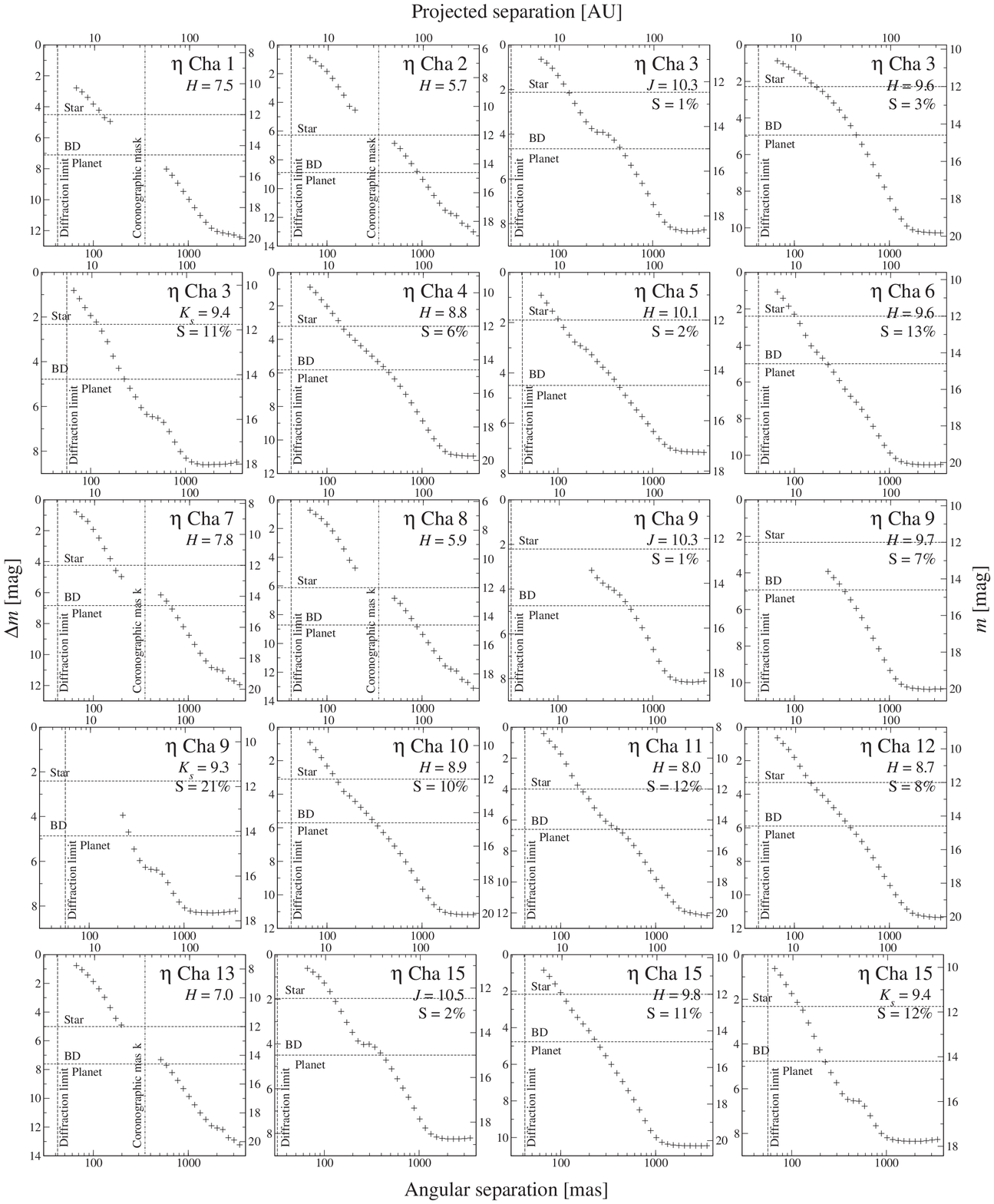}
\caption{\label{f:contrast1}See the caption of Fig.~\ref{f:contrast2}.}
\end{figure*}
\begin{figure*}
\includegraphics[width=\hsize,clip=true]{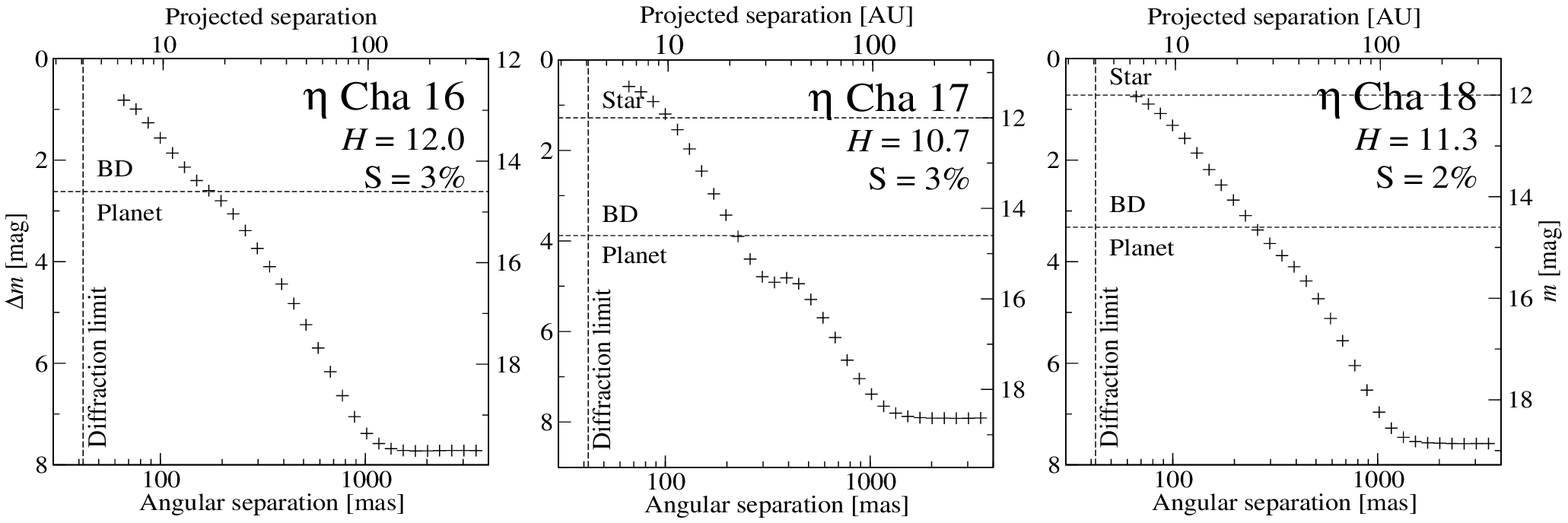}
\caption{\label{f:contrast2} Sensitivity to companions as a function of 
separation, determined as described in \S\,\ref{s:contrast}. The photometric
band and magnitude is given in each panel, as well as the measured Strehl ratio
(except for the coronographics observations, see \S\,\ref{s:strehl}). Left vertical axes
show the contrast sensitivity, while the right axes show the absolute
sensitivity. The bottom horizontal axes show the angular separation
and the upper axes the corresponding projected separation at the
$\eta$\,Cha cluster distance of 97\,pc. The diffraction limit and
the semitransparent coronographic mask radius are shown as vertical 
dashed lines. The two horizontal dashed lines depict the star/brown
dwarf (0.08\,M$_{\sun}$: $J=12.5$, $H=12.0$, and $K=11.8$) and brown dwarf/planetary 
mass (15\,M$_{\mathrm{J}}$: $J=15.0$, $H=14.6$, and $K=14.2$) boundaries for the \citet{bar03} 
evolutionary models at age 8\,Myr and distance 97\,pc.}
\end{figure*}
In order to estimate the sensitivity to companions as a function
of separation, we used the following procedure:
\begin{enumerate}
\item For every pixel $j$ in the detector, compute the distance $r_j$ to the
      determined center of the primary.
\item \label{en:affine} Choose two radii $R_0$ and $R_1$ and fit an affine function 
      $f(r) = a_0 + a_1r$, where $a_0$ and $a_1$ are fitting constants,
      to the intensity $I_j$ of all pixels $j$ with $r_j \in [R_0,R_1]$.
      Reject pixels more than 3$\sigma$ from the fit.
\item Compute the standard deviation of the fit residuals, 
      $\sigma_{\mathrm{pix}}(R) = \mathrm{StdDev}[I_j - f(r_j)]$, where
      $R$ is the mean of all $r_j \in [R_0,R_1]$.
\item Measure the FWHM of the primary PSF, and the number
      of pixels $N_{\mathrm{PSF}}$ and integrated flux $F_{\mathrm{PSF}}$
      within that area.
\item The derived 5$\sigma$ contrast sensitivity is then estimated to be
\begin{equation}
 \Sigma(R) = -2.5 \log_{10}\left[ \frac{5\sigma_{\mathrm{pix}}(R)
  \sqrt{N_{\mathrm{PSF}}}}{F_{\mathrm{PSF}}}\right]\,\mathrm{mag}.
\end{equation}
\end{enumerate}
The \textit{absolute} sensitivity is obtained by adding the primary magnitude
to $\Sigma(R)$. The function $f(r)$ in step~\ref{en:affine} is fit to remove
the radial gradient in the pixel intensities, that is due to the PSF. We confirmed 
that this procedure accurately estimates our achieved sensitivity by artificially
placing intensity scaled PSFs at various separations. Since we limit the
search to within 5\arcsec, we do not correct for anisoplanatism (see \S\,\ref{s:astrophot}).

For the semi-transparent coronographic observations, we made use of the 
measured suppression ratio in $H$ ($6.0\pm0.1$~mag, \S\,\ref{s:obs}).

In the case of $\eta$\,Cha\,9, which is a tight 0\farcs2
binary, we computed the contrast sensitivity from the center of
light of the two stars, and estimated $N_{\mathrm{PSF}}$ and 
$F_{\mathrm{PSF}}$ as the average and sum, respectively, of
both stars. The iso-intensity curves from the combined pair 
are elliptical rather than circular symmetric close to the
stars, which is the reason the contrast sensitivity is estimated
only at $\ge0\farcs2$ from the common center.

The contrast sensitivities for all observations are presented in 
Figs.~\ref{f:contrast1}\,\&\,\ref{f:contrast2}.


\subsection{Spectra of companion candidates}
\label{s:spectra}
\begin{figure}
\includegraphics[width=\hsize,clip=true]{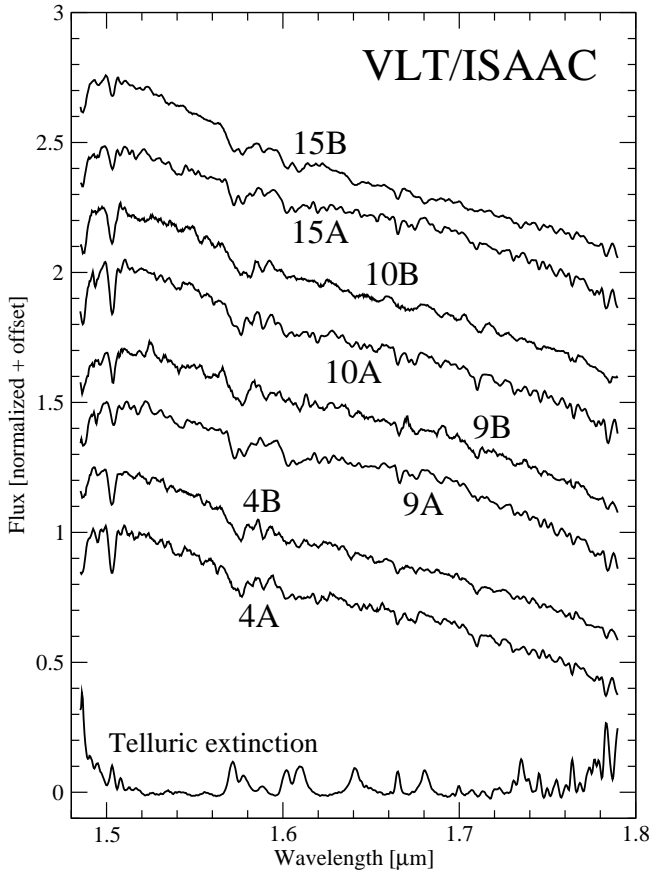}
\epsscale{0.8}
\caption{\label{f:isaac}ISAAC spectra of the primaries and companion
candidates of $\eta$\,Cha 4, 9, 10, and 15, denoted in the figure. 
The spectra have been normalized and offset in steps of 0.25. The
lowest spectrum shows the telluric atmospheric extinction, also
normalized and offset to 0. All of these companion candidates are
likely background objects (\S\,\ref{s:companions}).}
\end{figure}

\begin{figure}
\includegraphics[width=\hsize,clip=true]{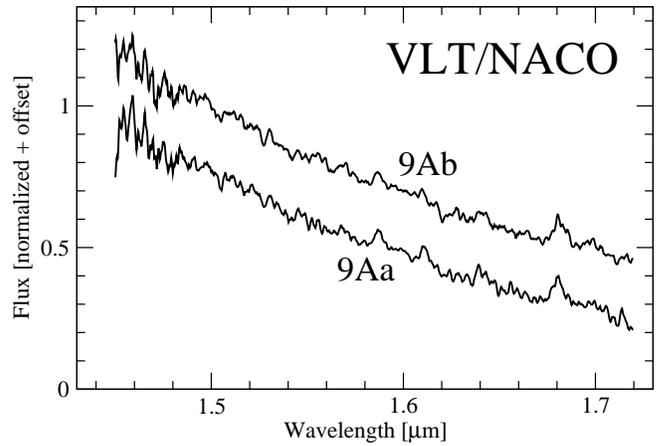}
\epsscale{1.0}
\caption{\label{f:nacospec} Resolved NACO spectra of the inner 
0\farcs2 $\eta$\,Cha~9 binary, normalized and offset by 0.25
from each other. The spectra are virtually identical. The scale
has been chosen so that the flux scale / wavelength scale ratio is the same as
in Fig.~\ref{f:isaac}.}
\end{figure}
The obtained ISAAC $H$-band spectra of the primaries and companion
candidates of the systems $\eta$\,Cha 4, 9, 10 and 15
are shown in Fig.~\ref{f:isaac}. Since the correction for telluric
lines was poor due to high airmass, a plot of the extinction is
also shown in Fig.~\ref{f:isaac}. 
The NACO $H$-band spectrum of the inner 0\farcs2 binary of $\eta$\,Cha~9 
is shown in Fig.~\ref{f:nacospec}.

\section{Discussion}
\label{s:discuss}
%

\subsection{Companion candidates}
\label{s:companions}

The $\eta$\,Cha cluster is only 22\degr\ from the Galactic plane,
so the probability of chance alignment stars in the 
13\farcs6$\times$13\farcs6 field of view is non-negligible. From the 2MASS 
All-Sky Data Release we find that the density of stars brighter than $H\sim16$
is $\sim10^{-3}$\,arcsec$^{-2}$, that statistically would produce 
about $3 \pm 2$  chance alignments in our 17 fields, consistent with the total 
number of companion candidates (5) with $H<16$ in our sample. Since
the observations are generally much more sensitive than this, we expect
even more faint background stars. We have unfortunately no access to
a deeper NIR survey of the region to find the local star
density, but a general conclusion is that potentially \textit{all} of
the 10 new companion candidates found (Table~\ref{t:bg}) could be chance 
alignments. 

For four of the candidates (near $\eta$\,Cha 4, 9, 10, and 6\farcs4 from $\eta$\,Cha~15) 
we have NIR ISAAC $H$-band spectra. If these companion candidates had been members of 
the $\eta$\,Cha cluster, their age ($\sim$8\,Myr) and 
luminosity (from Table~\ref{t:bg} and the distance 97\,pc) would have implied 
very low-mass objects with atmospheres cooler than 2500\,K \citep{bar03},
corresponding to spectral types later than M8.5 \citep{kir99,luh03}.
This is clearly not the case,
as the ISAAC spectra reveal all companion candidates to
have spectral types equal to, or earlier than, their primaries
(Fig.~\ref{f:isaac}). In particular, there is no evidence for water
depression that should be visible at $\sim$\,1.5$\,\mu$m for spectral 
types later than M5 \citep{cus05}. We thus conclude that these companion 
candidates most likely are background stars, and not associated 
with the $\eta$\,Cha cluster.

For the companion candidate near $\eta$\,Cha~3, we have two epochs of data separated
by 1.25\,yr that show a relative position difference of 
$(\Delta\alpha,\Delta\delta)$ = $(48\pm7,-14\pm7)$\,mas, implying a relative proper motion
of $(\mu_{\alpha},\mu_{\delta}) = (38\pm6,-11\pm6)$\,mas\,yr$^{-1}$. A common proper 
motion is thus ruled out with $>5\sigma$-significance, while the relative proper motion 
is roughly consistent with the companion
candidate being a background star, since the proper motion of the $\eta$\,Cha cluster
is $(\mu_{\alpha},\mu_{\delta}) = (-30.0\pm0.3,27.8\pm0.3)$\,mas\,yr$^{-1}$ \citep{mam00}.

The inner companion candidate of $\eta$\,Cha~3 also has two epochs of data, but unfortunately
the positional precision is not sufficient to significantly constrain the relative proper motion.
The non-membership status of the star is instead revealed by its NIR colors; a $H=17.5$ $\eta$\,Cha
member should have had $J-K \sim 3$ rather than $J-K = 0.7\pm0.7$.

The companion near candidate $\eta$\,Cha~1 is separated and bright enough to have 2MASS and DENIS photometry
(Table~\ref{t:bg}). We reject this as a physical companion, since the observed $I-(J,H,K)$ 
colors are $\sim$1\,mag fainter than expected from the \citet{bar98} models, even when allowing for 
variability due to the epoch difference between 2MASS and DENIS.

The remaining 3 companion candidate stars (near $\eta$\,Cha 2, 7, and 8)
have only single epoch $H$-band imaging, and therefore no direct way of 
ruling them out as physical companions. They are all quite distant from the system
primary ($>5.6\arcsec \sim 540$\,AU) and faint ($H\ga16$), however, and therefore 
likely chance alignments.

We conclude that, among the 17 $\eta$\,Cha members surveyed, there are no detected companions with 
projected separations 20--500\,AU.


\subsection{Orbit of $\eta$\,Cha~1}
\label{s:orbit}
\begin{figure}
\center{\includegraphics[width=\hsize,clip=true]{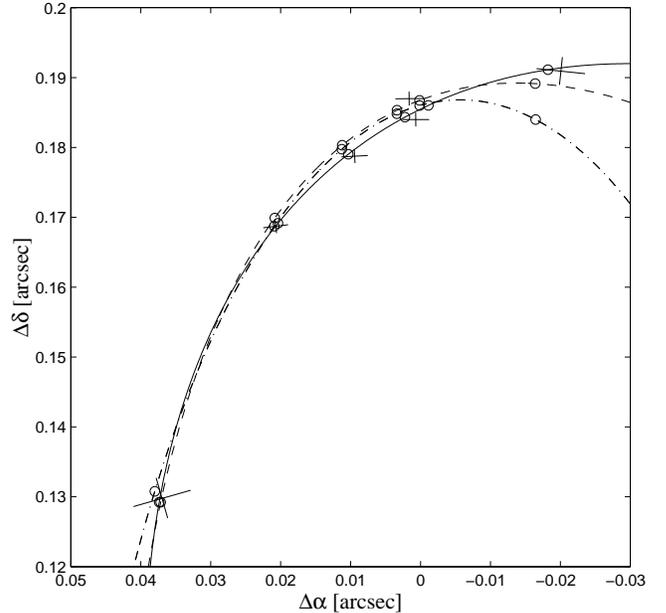}}
\caption{\label{f:orbit}Relative astrometry between the two components of 
$\eta$\,Cha~1. The crosses show the measured separations, with the 
right-most data point coming from this paper and the others from \citet{koh02}. The 
1$\sigma$ positional errors are indicated by the size of the crosses.
The dashed and dash-dotted curves are the orbit solutions found by 
\citet[see their Table~4]{koh02}, with periods of 150.8\,yr and 42.5\,yr, 
respectively. The unbroken line is the unphysical 93\,yr period orbit
fit mentioned in \S\,\ref{s:orbit}. The small circles denote the positions in the orbits 
that correspond to the dates of the observations.
}
\end{figure}

\citet{koh02} used multi-epoch observations of $\eta$\,Cha~1 between
1996-03-29 and 2001-12-10 to compute a preliminary orbit, estimating
a dynamical mass. With the additional data point from 2004-04-03, we
can constrain the orbit further (see Fig.~\ref{f:orbit}). In particular, among
the two preferred orbits with periods of 43\,yr and 151\,yr found in
\citet{koh02}, the longer orbit is clearly favoured by our data. To possibly
find a better orbital solution, we developed a simple orbit-fitting code that
works as follows:
\begin{enumerate}
\item For a given orbit $k$, compute the positions $(s^{k}_i,\phi^{k}_i)$ 
  on sky for the dates of the observations, where $s^{k}_i$ is the separation 
  of orbit $k$ and observation $i$, and $\phi^{k}_i$ the corresponding position angle. 
\item Compute the square sum  $\chi^2$ of the differences between the positions 
      $(s^{k}_i,\phi^{k}_i)$ from the assumed orbit, and the observed positions
      $(s^{\mathrm{obs}}_i,\phi^{\mathrm{obs}}_i)$:
\begin{displaymath}
\chi^2(k) = \sum_{i=1}^N \left[ \left(\frac{s^{k}_i - s^{\mathrm{obs}}_i}{\sigma_{s,i}}\right)^2
          + \left(\frac{\phi^{k}_i - \phi^{\mathrm{obs}}_i}{\sigma_{\phi,i}}\right)^2    \right],
\end{displaymath}
where $\sigma_{s,i}$ is the error in separation and $\sigma_{\phi,i}$ the error in 
position angle (including estimated systematic errors). 
\item Find the orbit $k$ that minimizes $\chi^2(k)$.
\end{enumerate}

The reason for using polar coordinates is that the position angle normally 
introduces a larger error than the separation, due to uncertainties in detector orientation. 

While the 151\,yr orbit of \citet{koh02} produces a good fit (Fig.~\ref{f:orbit}), we find a multitude of 
very different orbits that produce equally good or better fits. In particular,
there is a general solution degeneracy such that extremely eccentric orbits
at high inclinations produce good fits (but with an unrealisticly high system
mass). An example is given in Fig.~\ref{f:orbit}, where the unbroken line shows an orbit
with a period of 93\,yr and semi-major axis 0\farcs416, corresponding to the
system mass of 7.6\,M$_{\sun}$ at the distance of 97\,pc. Our conclusion is 
that the orbit has to be followed for a longer time span before useful limits
on the dynamical mass can be made. Alternatively, spatially resolved radial 
velocity measurements would add valuable constraints.


\subsection{Limits on companion probability}
\label{s:limits}
The contrast sensitivity estimates from \S\,\ref{s:contrast} can
be used to put limits on the number of likely companions. The basic approach
is to assume that companions of stars are assigned by a stochastic 
process such that any given star system will have a companion with 
probability $p$, called the multiplicity (or binary) probability. 
The observed systems are then seen as a sample of
this stochastic variable. Even if we knew with complete certainty
that the 17 systems of $\eta$\,Cha had no wide stellar companions, there
would still be a 5\% chance that this outcome would have been produced
with $p = 0.16$; thus, the 95\% confidence upper limit for $p$
would have been $0.16$.

Before we can do the proper statistics, however, we need to correct
for the observational biases. That is, given that there is a companion, 
would we have detected it? 
Let $p$ be the probability that a star has a companion, and $q_j$ the
probability that a companion would have been detected in system $j$, given 
that there had indeed been a companion there. Then the probability
that a companion is not detected in system $j$ is $1 - p q_j$, and the
probability that no companions are detected in $N$ systems enumerated 
from 1 to $N$ is
\begin{equation}
\label{e:upper}
\Phi(p) = \prod_{j=1}^N (1 - p q_j).
\end{equation}

With a confidence set we mean the set of $p$ such that 
$\Phi(p) \le 1 - \alpha$, where $\alpha$ is the chosen confidence, 
typically $\alpha = 95\%$. As $\Phi(p)$ is a monotonously decreasing 
function of $p$, the confidence set becomes an interval 
$[0,p_{\alpha}]$, where we call $p_{\alpha}$ the $\alpha$-confidence
upper limit on $p$.


\begin{figure*}
\includegraphics[width=\hsize,clip=true]{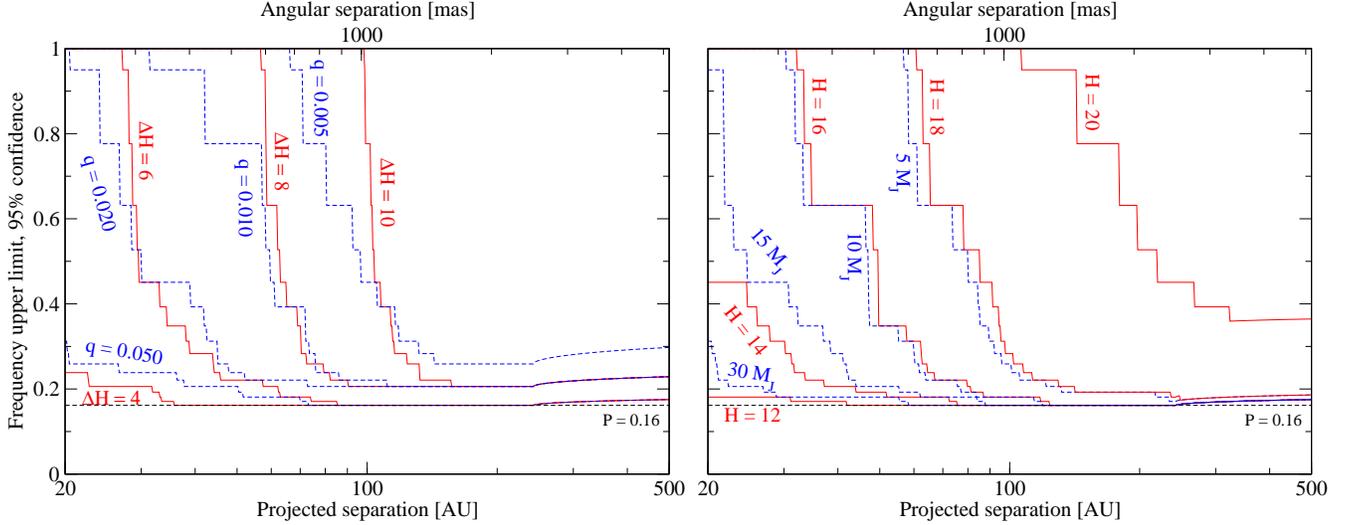}
\caption{\label{f:sep}
The 95\% confidence upper limits on the probability
of companions. The left panel shows the upper limits for companions of
specific contrasts to the primary, as a function of separation from the 
star. The dashed lines (colored blue in electronic edition) show the
sensitivity limits to mass ratios $q = M_{\mathrm{comp}}/M_{\mathrm{prim}}$,
while the unbroken lines (colored red) show the limits for flux-ratios
$\Delta H$. The right panel displays the absolute sensitivity,
where the dashed lines (colored blue) show sensitivity to companion
masses, and the unbroken lines (colored red) show the sensitivity to
apparent $H$-band magnitudes. The dashed horizontal line 
shows the 95\% confidence upper limit that would have been obtained statistically
for our sample of 17 systems, had the detection sensitivity been 100\%. The
increased upper limits at separations $>$2\farcs5 is due to part of the field
being outside the array.
}
\end{figure*}

To derive $p_{\alpha}$ we need to know $q_j$ for all observed systems.
For a given magnitude and separation from the star, this becomes straightforward 
using the contrast sensitivity estimates from  \S\,\ref{s:contrast}. If the 
brightness is above the sensitivity limit at the 
given separation in system $j$, $q_j = 1$; else $q_j = 0$.  
Equation~\ref{e:upper} thus reduces to $\Phi(p) = (1 - p)^n$, where $n$ is the
number of systems where detection would have been possible. This implies
$p_{\alpha} = 1 - (1 - \alpha)^{1/n}$. Because $p_{\alpha}$ thus only 
depends on the discrete $n$, $p_{\alpha}$ will also be discrete, as
shown in Fig.~\ref{f:sep}. 
In the left panel of Fig.~\ref{f:sep} we derive 95\% upper 
limits on the probability of
companions of specific flux ratios to the primary, as a function of
separation. To derive limits on the companion probability for specific 
mass ratios $q$, we assume primary masses from \citet{lyo04b}, and use
evolutionary models from \citet{bar03} to translate companion mass to an
$H$-band magnitude, assuming the age 8\,Myr, distance 97\,pc and solar metallicity.
In the right panel of Fig.~\ref{f:sep} we instead display limits on the absolute 
sensitivity to companions of different $H$-band magnitudes and masses.

From a physical point of view, rather than knowing the companion probability
as a function of observed (projected) separation, it is more interesting to 
find the probability as a function of semi-major axis. To do that we need to
know the probability $q_j$ to find a companion, given that it has a specific
semi-major axis. Our ability to detect a companion will depend on its
projected distance to the primary, which in turn depends on viewing
geometry and orbital phase (for eccentric orbits). In the Appendix, we
compute the projected separation probability distribution for a 
companion of semi-major axis $a$. Assuming a random orientation, we derive 
the distribution 
analytically for circular orbits (\S\,\ref{s:circular}), and use a numerical 
approach in the case of eccentric orbits with an eccentricity distribution 
$f(e)=2e$ (motivated by both theory and observations; see 
\S\,\ref{s:elliptical}). The probability $q_j$ is then 
the probability that the companion is located at 
a detectable distance from the binary. Assuming that the sensitivity increases
monotonously as a function of separation, this equals the 
probability that the companion is outside the ``detection separation'' $d$
-- i.e.\ 
\begin{equation}
\label{e:qj}
q_j = \int_{d/a}^{\infty} f_S(s)\,\mathrm{d}s = 1 - F_S(d/a),
\end{equation}
where $f_S(s)$ is the projected separation probability density distribution 
found in the Appendix and shown in Fig.~\ref{f:prob}, and  $F_S(s)$ is the corresponding
probability distribution. Since the projected 
separation can be arbitrary small for any semi-major axis, we always have
that $F_S(d/a) > 0$ for $d/a > 0$, and thus $q_j < 1$. Once we know $q_j$, 
equation~\ref{e:upper} is used to compute 95\% upper limits on the probability 
of companions. For the coronographic observations, the mask is at a fixed
position only $h=2\farcs5$ from the edge of the detector; we therefore introduce
the additional correction factor $g(s) = 1 - \pi^{-1}\arccos(h/s)$ (for
$s > h$) into the integral of equation~\ref{e:qj}, where $g(s)$ is the probability that a
companion at separation $s$ is in the field of view.


\begin{figure*}
\includegraphics[width=\hsize,clip=true]{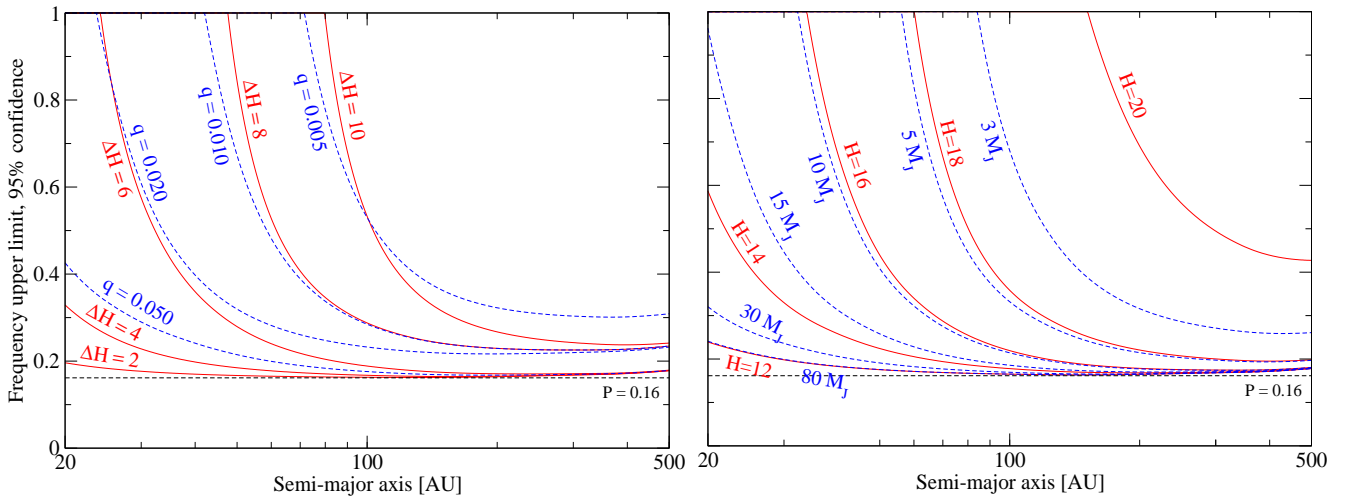}
\caption{\label{f:major} 
Same as in Fig.~\ref{f:sep}, except that the sensitivity limits are displayed
as a function of semi-major axes, assuming the eccentricity
probability density distribution $f(e) = 2e$.}
\end{figure*}

In Fig.~\ref{f:major} we show the 95\% upper limits in the same way as in Fig.~\ref{f:sep},
but as a function of the semi-major axes.


\subsection{Deficit of wide binaries}
\label{s:deficiency}
It is remarkable that, despite our high sensitivity, the \emph{widest} binary 
detected in the $\eta$\,Cha cluster is only
0\farcs21, corresponding to the projected distance of 20\,AU. This lack of wide binaries
was already noted by \cite{koh02}, albeit with smaller statistics and less sensitive measurements.
Outside 20\,AU, we would have found any stellar
companion (of mass $>$0.07\,M$_{\odot}$; see Fig.~\ref{f:sep}). The fact that
we do not detect any wide binaries among the 17 members implies that
the wide ($>30$\,AU) binary probability $w_{\eta\,\mathrm{Cha}}$ with 1$\sigma$
confidence is lower than $p_{1\sigma}$=0.07, and with 95\% confidence is lower
than $p_{95\%}$=0.18 (Fig.~\ref{f:major}). This stands in stark contrast to the other 
nearby, young (of the similar age 8\,Myr) TW~Hydrae association (TWA), 
where 11 out of the member stars TWA~1--19 are
binaries with separations $\ga 30$\,AU \citep{bra03}, corresponding to a wide 
binary probability
of $w_{\mathrm{TWA}}=0.58^{+0.13}_{-0.14}$, clearly different from what
we find in $\eta$\,Cha. 
This difference cannot be due to detection sensitivity differences; all the
companions in TWA would have easily been detected in the $\eta$\,Cha cluster 
by this survey. The difference cannot be explained by small number statistics
either, as the statistical likelihood that the $\eta$\,Cha cluster
and TWA have binary frequencies drawn from the same distribution is less than 
$2\times10^{-4}$ (see the Appendix). 

There is certainly no deficiency of \emph{close} binaries ($< 30$\,AU) in the
$\eta$\,Cha cluster \citep{lyo04b}. If anything, there might be a small 
over-abundance, compared to TWA, but the statistics are inconclusive.

One may speculate on the cause for such a difference in the wide binary
population between the $\eta$\,Cha cluster and the TWA. One notable difference is that 
the $\eta$\,Cha cluster is much denser than the TWA. Could it be that wide binaries are 
dynamically disrupted in the $\eta$\,Cha cluster? To investigate this possibility we
estimate the timescale for a binary to undergo a strong encounter with another
star in the cluster. From \citet[equation~7]{iva05}, we have that this timescale is
\begin{eqnarray}
\label{e:colltime}
\tau_{\mathrm{coll}} = 1.7 \times 10^{8}\,\mathrm{yr} \times
\eta^2 k^{-2} n_5^{-1}\frac{\langle M \rangle^2}{M_1^2M_2^2} \nonumber\\
\times \left(1 + \eta \frac{2}{k}\frac{M_1 + M_2 + 
\langle M \rangle}{M_1 M_2} \langle M \rangle \right)^{-1},
\end{eqnarray}
where $\eta$ is the hardness of a binary, $k \simeq 2$, $n_5$ is the number
density of star systems in units of $10^5$\,pc$^{-3}$, $\langle M \rangle$
is the mean mass per star system in units of M$_{\sun}$, and $M_1$ and $M_2$ are 
the masses of the binary components, also in units of M$_{\sun}$. The hardness 
of a binary is defined as
\begin{equation}
\eta = \frac{M_1 M_2}{\langle M \rangle \sigma^2 a} G \mathrm{M}_{\sun},
\end{equation}
where $G$ is the gravitational constant, $a$ is the binary separation, 
and $\sigma$ is the velocity dispersion 
of the cluster. Binaries with $\eta < 1$ are termed soft, and those with 
$\eta > 1$ hard. In general, soft binaries are disrupted by strong encounters, 
while hard binaries may survive. In the $\eta$\,Cha cluster, we have the cumulative
mass 16.6\,M$_{\sun}$ distributed over 17 systems \citep{lyo04b}, 
giving $\langle M \rangle = 1.0$. 
There are presently no accurate radial velocities published for the $\eta$\,Cha 
members. However, from the estimated age of the cluster ($\sim$8\,Myr) and
observed effective radius ($\sim0.2$\,pc), we infer that the cluster most likely is
gravitationally bound. The present escape velocity is namely 
$v_{\mathrm{esc}} = (2GM/R)^{1/2} \sim 0.75$\,km\,s$^{-1}$, assuming $M$ = 13\,M$_{\sun}$
within $R = 0.2$\,pc \citep{lyo04b}, which gives a crossing time of merely 0.3\,Myr,
enough to traverse the cluster core 25 times during its lifetime, while the relaxation
time is only a few times the crossing time \citep[chapter 4]{bin87}. We 
therefore assume  $\sigma = v_{\mathrm{esc}}/2 = 0.37$\,km\,s$^{-1}$, in accordance
with the virial theorem, implying that the condition for a binary to
be soft in the $\eta$\,Cha cluster is $a > 1600$\,AU $\gg 30$\,AU for the typical masses 
$M_1 = M_2 = 0.5$. 
Moreover, for the $\eta$\,Cha cluster $n_5 \sim 3\times10^{-3}$, and the collision timescale 
for $\eta=1$ binaries is thus $\tau_{\mathrm{coll}}\sim26$\,Gyr, i.e.\ 3000 times 
longer than the lifetime of the system. That dynamical interactions between binaries
in the present configuration of the cluster should be responsible for the lack of 
wide ($>$30\,AU) binaries, consequently seems highly unlikely. 

Since both the $\eta$\,Cha cluster and TWA are of similar age, the remaining 
explanation is that the difference in multiplicity properties were imprinted during 
the formation phase, as a result of different initial conditions. Either the groups
formed with different multiplicity properties, or the properties 
dynamically evolved very early on, when the stars were possibly much closer together.

There seems to be a general trend that denser groups have smaller wide binary
frequencies than sparser regions; the sparse regions Taurus, Ophiuchus and 
Chamaeleon \citep{duc99}, and MBM~12 \citep{bra03} all have high wide binary
frequencies, while the denser regions Trapezium \citep{pet98}, and NGC~2024, 2068, and 
2071 \citep{pad97}, have low wide binary frequencies. 
Solar-type main-sequence stars
in the solar neighborhood have a binary probability of $\sim$0.45 
\citep[and references therein]{lei93}, with half being wide 
\citep{duq91}, and thus $w_{\mathrm{MS}}=0.23$, which is right in between 
$w_{\eta\,\mathrm{Cha}}$ and $w_{\mathrm{TWA}}$.
It is therefore not clear from the wide binary statistics alone if sparse
or dense star formation is the dominant mode -- possibly both contribute 
equally. What is clear is that models of cluster formation and early evolution 
probably are essential to explain multiplicity properties 
\citep[e.g.][]{bat02,del04,goo06}.

\section{Conclusions}
\label{s:conclusions}
%
%
We summarize our conclusions as follows:

\begin{enumerate}
\item We found no new companions to stars in the $\eta$\,Cha
      cluster, despite being sensitive down to the star/BD limit outside
      0\farcs3 (30\,AU) and down to the BD/planet limit outside  0\farcs5
      (50\,AU).

\item We have constrained the orbit of $\eta$\,Cha\,1 further, but are unable
      to usefully constrain a dynamical mass. Resolved radial velocities of the
      two components, or a longer astrometric time baseline, are required for an 
      accurate mass estimate.

\item The 95\,\% upper limit for the wide ($>$30\,AU) binary probability $w_{\eta\,\mathrm{Cha}}$
      in $\eta$\,Cha is $p_{95\%}$=0.18. This contrasts to the wide binary
      probability $w_{\mathrm{TWA}}=0.58^{+0.13}_{-0.14}$ of TWA. The
      likelihood that  $w_{\eta\,\mathrm{Cha}} = w_{\mathrm{TWA}}$ is less than
      $2\times10^{-4}$.

\item Multiplicity properties depend on the initial conditions of the 
      formation environment.

\end{enumerate}

\acknowledgments

We acknowledge the outstanding support by the ESO user support department, 
in particular Sabine Mengel and Danuta Dobrzycka. We also thank Nancy Ageorges 
for help in determining the contrast of the semi-transparent mask.
We are grateful to Marten van Kerkwijk, Natasha Ivanova and Serge Correia for 
valuable discussion.
This research was supported by an NSERC grant and an SAO subcontract for the 
Keck Nuller project to RJ.
We made extensive use of NASA's Astrophysics Data System
Bibliographic Services, and the SIMBAD database and VizieR catalogue access tool,
operated at CDS, Strasbourg, France. We used
data products from the Two Micron All Sky Survey and the DENIS consortium.

\textit{Facilities:} \facility{VLT:Yepun (NACO)}, \facility{VLT:Antu (ISAAC)}.

\begin{appendix}


\section{Probability distributions for projected binary separations}

This appendix concerns the computation of probability distributions
of the observed projected separation between the two components of a
binary system, given that the semi-major axis is known. Here we use
the semi-major axis $a$ of the companion orbit \emph{relative to the primary}.
If the mass of the companion is a significant fraction of the primary, 
this will be different from the semi-major axis $a_{\mathrm{com}}$ of 
the companion orbit relative to the center of mass. The relation is
$a = (M_{\mathrm{comp}} / M_{\mathrm{prim}} + 1) a_{\mathrm{com}}$,
where $M_{\mathrm{comp}}$ and $M_{\mathrm{prim}}$ are the masses of the
companion and primary, respectively. The derived probability distributions
below are easily scaled to $a_{\mathrm{com}}$ in case of specific 
mass-ratio systems.

\subsection{Companions in circular orbits}
\label{s:circular}

To compute the 
probability distribution for the projected separation of a companion, we
need to make the assumption that the orbit is observed from a uniformly
distributed direction; that is, any viewing direction is equally probable.

To generate a stochastic vector with a direction
uniformly distributed over the unit sphere is a less trivial problem
than one might naively expect. One cannot simply use a uniform distribution of
spherical coordinates, since this would bias the vectors too much towards the
poles; and one cannot simply use uniformly distributed Cartesian coordinates,
since this would bias the directions towards the corners. Instead, a stochastic 
vector $\mathbf{d}=(x,y,z)$ uniformly distributed over the unit sphere may be 
generated as
\begin{eqnarray*}
x &=& \sqrt{1-z^2}\cos\phi \\
y &=& \sqrt{1-z^2}\sin\phi \\
z &\in& U(-1,1),
\end{eqnarray*}
where $X \in U(a,b)$ means that $X$ is a stochastic variable uniformly 
distributed between $a$ and $b$, and $\phi \in U(0,2\pi)$. This is a corollary
from a theorem by Archimedes, that states that a lateral area of a section cut 
out of a sphere by two parallel planes equals $A = 2 \pi R h$, where
$R$ is the radius of the sphere and $h$ is the distance between the planes.

With $\mathbf{R}$ being the radius vector from the star to the companion,
$R$ its length and $\mathbf{d}$ the unit
viewing direction vector, the projected distance between the star and the
planet onto a plane perpendicular to the viewing direction will be
$s = \sqrt{R^2 - (\mathbf{R \bullet d})^2}$.

Since $\mathbf{d}$ is uniformly distributed on the unit sphere, we can (without
loss of generality) let $\mathbf{R} = R \mathbf{e}_\mathrm{z}$, where $\mathbf{e}_\mathrm{z}$ is the
unit vector along the z-axis. The scalar product becomes
$\mathbf{R \bullet d} = R d_\mathrm{z}$, where $d_\mathrm{z}$ is the z-component
of the direction vector $\mathbf{d}$. That is, the scalar product is only dependent
on the z-component of the direction vector:
\begin{equation}
\label{e:projection}
s = R \sqrt{1 - d_\mathrm{z}^2}.
\end{equation}
The probability distribution for the corresponding stochastic variable
$S = \sqrt{1 - D^2}$, where $D \in U(0,1)$, becomes $F_S(s) = 1 - \sqrt{1 - s^2}$, 
and the probability density distribution $f_S(s) = \mathrm{d}F_S(s)/\mathrm{d}s = s/\sqrt{1 - s^2}$.

\begin{figure}
\center{\includegraphics[width=\hsize,clip=true]{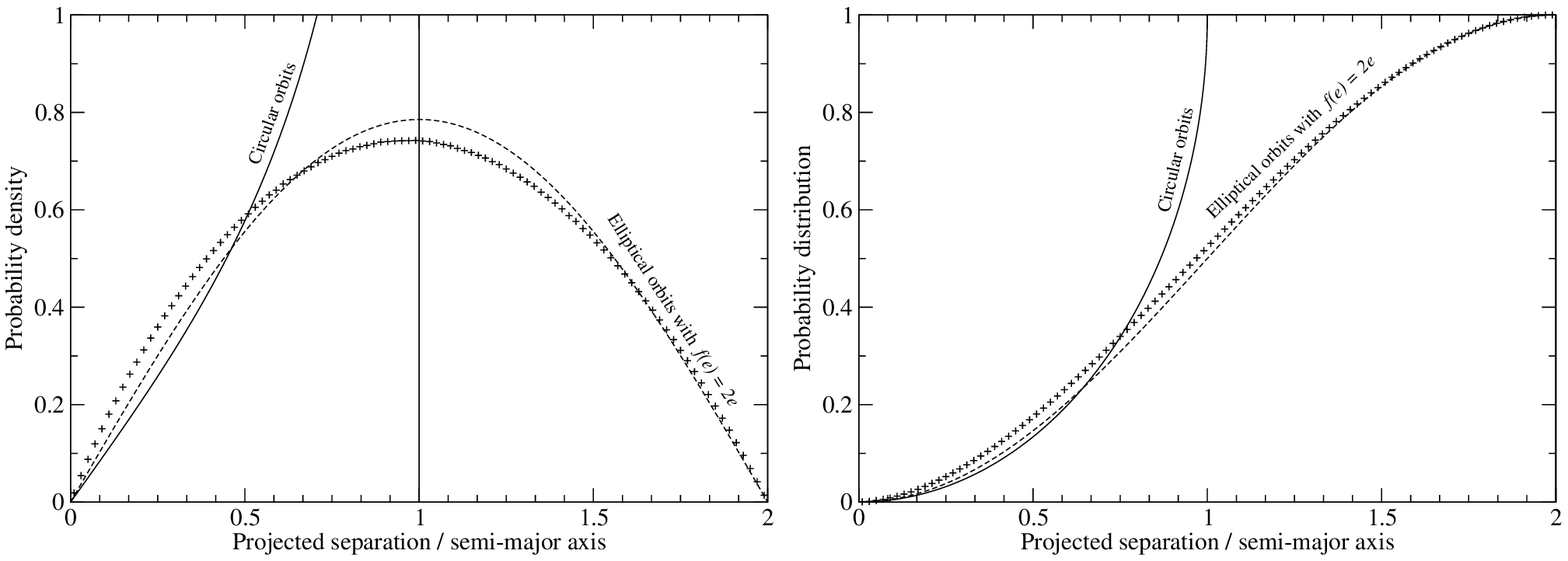}}
\caption{\label{f:prob}The left panel shows the probability density distribution 
for the projected separation of a companion to a star, in units of its semi-major axis. Two
cases are plotted, one where the orbits are assumed to be circular (solid line),
and one where the eccentricity density distribution is assumed to be $f(e)=2e$
(crosses). The projected separation distribution is computed exactly for
circular orbits, and by a monte carlo approach for the eccentric orbits. The
dashed line is an approximate fit, as outlined in \S\,\ref{s:elliptical}. The
right panel shows the corresponding probability distribution, i.e.\ the integral over the
probability density.
}
\end{figure}

These distributions thus describe the projected separation $s$, in units of the orbital radius, 
of a companion in a circular orbit around a star (Fig.~\ref{f:prob}).


\subsection{Companions in elliptical orbits}
\label{s:elliptical}

To derive the projected separation distribution for elliptical orbits is slightly more 
complicated than for circular motion, since the radial distance
between the star and its companion is non-linear in time. We also have to
make an additional assumption on the eccentricity distribution. We still do
not have to bother with the orbital elements concerning a specific orientation 
of an orbit however, since we already have assumed that the viewing direction is
random.
 
The mean anomaly $M$ of an elliptical orbit is defined to be linearly increasing 
with time \citep[Chapter 29]{mee91}. Its relation to the eccentric anomaly $E$, that is used
to compute the actual position of the companion, is defined by the Kepler equation
$E = M + e\sin E$, where $e$ is the eccentricity of the orbit. Since the
Kepler equation is transcendental it cannot be solved algebraically, which
complicates the analysis. We therefore derive the probability density distribution
using the following monte carlo approach instead.
\begin{enumerate}
\item Let the mean anomaly $M \in U(0,2\pi)$.
\item Get the eccentricity $e$ from some pre-defined distribution $f(e)$.
\item Solve Kepler's equation numerically and compute the
      eccentric anomaly $E(M)$.
\item Compute the instantaneous distance between star and companion,
      $R = 1 - e\cos E$.
\item Let the projection term $d_{\mathrm{z}} \in U(-1,1)$,
      in line with equation~\ref{e:projection}, and compute the projected
      separation $s = R \sqrt{1 - d_\mathrm{z}^2}$.
\end{enumerate}
The choice of eccentricity distribution is important for the resulting
probability density distribution. 
Theoretical considerations \citep{amb37} predict
the distribution $f(e) = 2e$, where $e \in [0,1]$, which is reasonably
confirmed by observations of long-period binaries 
\citep[$>$1000~days;][]{duq91}. To generate a stochastic variable with
$f(e) = 2e$, we let $e = X^{1/2}$, where $X \in U(0,1)$.
The result from a simulation of $10^8$ binaries is shown in Fig.~\ref{f:prob}. The
probability density distribution is well approximated, although not perfectly, 
by the function
\begin{equation}
f_S(s) = \frac{\pi}{4}\sin\left(\frac{\pi}{2}s\right),
\end{equation}
where $s \in [0,2]$, which is over plotted in Fig.~\ref{f:prob}. 
The corresponding probability distribution 
is
\begin{equation}
F_S(s) = \frac{1}{2} \left[1 - \cos\left(\frac{\pi}{2}s\right)\right].
\end{equation}
A stochastic variable $S$ with this distribution can be generated
from $X \in U(0,1)$ by setting $S = 2 \pi^{-1} \arccos(1-2X)$, that might prove
useful for future completeness studies.


\section{Multiplicity statistics}

Statistical estimates of multiplicity frequencies are often limited by small
sample numbers, where approximate methods using assumptions of Poisson or
normal statistics become insufficient. Instead, more accurate estimates can
be obtained by explicit use of binomial statistics. The multiplicities of $N$ 
systems of a stellar association are viewed as outcomes $x_i$ of a binomially 
distributed stochastic variable $X \in Bin(1,p)$, where $p$ is the multiplicity 
probability for any specific system. $x_i = 1$ if system $i$ is multiple, and $x_i = 0$ 
if not. The number of multiple systems is then $k = \sum_i x_i$, which in itself
can be seen as the outcome of a binomially distributed stochastic variable 
$K \in Bin(N,p)$.
The goal is to constrain $p$ from a given set $\{x_i\}$, assuming all $x_i$
are drawn from the same distribution. Here we summarize results from two
outstanding problems, how to constrain the multiplicity probability $p$ by a 
confidence interval, and how to decide if two sets $\{x_i\}$ and $\{y_i\}$ are
outcomes of the same binomial distribution.

\subsection{Binomial confidence intervals}

The well-known maximum-likelihood estimator of $p$ is $\hat{p} = \sum_i x_i/N$, but
to compute an accurate variance of this estimator is much more difficult. The most
widely used way of computing the standard deviation of $\hat{p}$ is by using the 
so-called \textit{Wald} method, that has the simple form 
$\sigma = (\hat{p}(1-\hat{p})/N)^{1/2}$. The Wald standard deviation 
produces very poor approximations whenever $p$ is close to 0 or 1, however, and
its usage is generally not recommended \citep{bro01}. The problem of computing
accurate confidence intervals for the binomial distribution is not a new one, 
and there exist a plenitude of literature on the subject (see \citealt{bro01} 
for a review of methods). A robust ``exact'' method guaranteed to produce intervals with
confidence of \textit{at least} $\alpha$ was proposed by \citet{clo34}. If $K \in Bin(N,p)$,
the probability that an outcome is $k$ is 
$P(k~\mathrm{of}~N) = {N\choose k} p^k(1-p)^{N-k}$, and the probability that $k < n \le N$ is
$P(k < n~\mathrm{of}~N) = \sum_{k=0}^{n-1} P(k~\mathrm{of}~N)$. For any given significance
$\alpha$ and specific outcome $k$, the Clopper-Pearson method consists of finding the interval
$(p_{\mathrm{min}},p_{\mathrm{max}})$ such that if $p \ge p_{\mathrm{min}}$, then 
$P(j < k~\mathrm{of}~N) \le (1-\alpha)/2$, and if $p \le p_{\mathrm{max}}$, then 
$P(j > k~\mathrm{of}~N) \le (1-\alpha)/2$. By solving for the equalities, we get
\begin{eqnarray}
 \sum_{j=0}^{k} {N\choose j} p_{\mathrm{min}}^j(1-p_{\mathrm{min}})^{N-j} &=& \frac{1-\alpha}{2} \label{e:pmin}\\
\sum_{j=0}^{k-1}{N\choose j} p_{\mathrm{max}}^j(1-p_{\mathrm{max}})^{N-j} &=& \frac{1 + \alpha}{2}, \label{e:pmax}
\end{eqnarray}
where we have used that $P(j > k~\mathrm{of}~N) = 1 - P(j < k+1~\mathrm{of}~N)$. The equations 
\ref{e:pmin} \& \ref{e:pmax} are normally best solved numerically.\footnote{An applet
to compute binomial confidence intervals is available at \url{http://statpages.org/confint.html}} 
In the degenerate case where
$k=0$ or $k=N$, the confidence interval becomes one-sided, and the equations 
\ref{e:pmin} \& \ref{e:pmax} are easily solved analytically 
(e.g., equation \ref{e:upper} in \S\,\ref{s:limits}).

\subsection{Comparing binomial distributions}

Let $k_A$ out of $N_A$ systems in association $A$ be multiple, and $k_B$ out of $N_B$ 
systems in association $B$. The question that naturally arises is, is the multiplicity 
probability $p_A$ of system $A$ similar to $p_B$ of system $B$? As in the case
of binomial confidence intervals, there are several tests available in the literature
(see \citealt{sto90} for a review). A conservative ``exact'' hypothesis test is
based on \citet{fis35}, where the hypothesis that $p_A = p_B$ is tested: Let 
$h(k_A,k_B,N_A,N_B) = {N_A\choose k_A} {N_B\choose k_B} / {N_A+N_B\choose k_A+k_B}$
and $I[\mathit{expression}]$ be the indicator function that is 1 if $\mathit{expression}$
is true, and 0 otherwise. Then the test function is
\begin{equation}
T = \sum_{x=\max(0,k_A+k_B-n_B)}^{\min(n_A,k_A+k_B)} h(x,k_A+k_B-x,n_A,n_B) \times I\left[h(x,k_A+k_B-x,n_A,n_B) \le 
h(k_A,k_B,n_A,n_B)\right],
\end{equation}
and the hypothesis is rejected if $T \le \alpha$, where $\alpha$ is the significance of the test. As an
example, if $k_A = 0$, $N_A = 17$, $k_B = 11$, and $N_B = 19$, then $T = 1.46\times10^{-4}$, which is
the quoted likelihood that the wide binary probability is equal in $\eta$\,Cha and TWA (\S\,\ref{s:deficiency}).

\end{appendix}

\bibliography{ms}

\begin{thebibliography}{44}
\expandafter\ifx\csname natexlab\endcsname\relax\def\natexlab#1{#1}\fi

\bibitem[{{Adams} \& {Myers}(2001)}]{ada01}
{Adams}, F.~C., \& {Myers}, P.~C. 2001, \apj, 553, 744

\bibitem[{{Ambartsumian}(1937)}]{amb37}
{Ambartsumian}, V.~A. 1937, Astron.\ Zh., 14, 207

\bibitem[{{Baraffe} {et~al.}(1998){Baraffe}, {Chabrier}, {Allard}, \&
  {Hauschildt}}]{bar98}
{Baraffe}, I., {Chabrier}, G., {Allard}, F., \& {Hauschildt}, P.~H. 1998, \aap,
  337, 403

\bibitem[{{Baraffe} {et~al.}(2003){Baraffe}, {Chabrier}, {Barman}, {Allard}, \&
  {Hauschildt}}]{bar03}
{Baraffe}, I., {Chabrier}, G., {Barman}, T.~S., {Allard}, F., \& {Hauschildt},
  P.~H. 2003, \aap, 402, 701

\bibitem[{{Bate} {et~al.}(2002){Bate}, {Bonnell}, \& {Bromm}}]{bat02}
{Bate}, M.~R., {Bonnell}, I.~A., \& {Bromm}, V. 2002, \mnras, 336, 705

\bibitem[{{Binney} \& {Tremaine}(1987)}]{bin87}
{Binney}, J., \& {Tremaine}, S. 1987, {Galactic dynamics} (Princeton, NJ,
  Princeton University Press, 1987, 747 p.)

\bibitem[{{Brandeker} {et~al.}(2003){Brandeker}, {Jayawardhana}, \&
  {Najita}}]{bra03}
{Brandeker}, A., {Jayawardhana}, R., \& {Najita}, J. 2003, \aj, 126, 2009

\bibitem[{{Brown} {et~al.}(2001){Brown}, {Cai}, \& {DasGupta}}]{bro01}
{Brown}, L.~D., {Cai}, T.~T., \& {DasGupta}, A. 2001, Statistical Science, 16,
  101

\bibitem[{{Clopper} \& {Pearson}(1934)}]{clo34}
{Clopper}, C.~J., \& {Pearson}, E.~S. 1934, Biometrika, 26, 404

\bibitem[{{Cushing} {et~al.}(2005){Cushing}, {Rayner}, \& {Vacca}}]{cus05}
{Cushing}, M.~C., {Rayner}, J.~T., \& {Vacca}, W.~D. 2005, \apj, 623, 1115

\bibitem[{{Delgado-Donate} {et~al.}(2004){Delgado-Donate}, {Clarke}, {Bate}, \&
  {Hodgkin}}]{del04}
{Delgado-Donate}, E.~J., {Clarke}, C.~J., {Bate}, M.~R., \& {Hodgkin}, S.~T.
  2004, \mnras, 351, 617

\bibitem[{{Duch{\^e}ne}(1999)}]{duc99}
{Duch{\^e}ne}, G. 1999, \aap, 341, 547

\bibitem[{{Duch{\^e}ne} {et~al.}(2006){Duch{\^e}ne}, {Delgado-Donate},
  {Haisch}, {Loinard}, \& {Rodriguez}}]{duc06}
{Duch{\^e}ne}, G., {Delgado-Donate}, E., {Haisch}, K., {Loinard}, L., \&
  {Rodriguez}, L. 2006, Protostars and Planets V, preprint available at
  astro-ph/0603004

\bibitem[{{Duquennoy} \& {Mayor}(1991)}]{duq91}
{Duquennoy}, A., \& {Mayor}, M. 1991, \aap, 248, 485

\bibitem[{{Fisher}(1935)}]{fis35}
{Fisher}, R.~A. 1935, Journal of the Royal Statistical Society, 25, 295

\bibitem[{{Goodwin} {et~al.}(2006){Goodwin}, {Whitworth}, \&
  {Ward-Thompson}}]{goo06}
{Goodwin}, S.~P., {Whitworth}, A.~P., \& {Ward-Thompson}, D. 2006, \aap, 452,
  487

\bibitem[{{Haisch} {et~al.}(2005){Haisch}, {Jayawardhana}, \& {Alves}}]{hai05}
{Haisch}, K.~E., {Jayawardhana}, R., \& {Alves}, J. 2005, \apjl, 627, L57

\bibitem[{{Houk} \& {Cowley}(1975)}]{hou75}
{Houk}, N., \& {Cowley}, A.~P. 1975, {Michigan Catalogue of two-dimensional
  spectral types for the HD star} (Ann Arbor: University of Michigan,
  Departement of Astronomy, 1975)

\bibitem[{{Ivanova} {et~al.}(2005){Ivanova}, {Belczynski}, {Fregeau}, \&
  {Rasio}}]{iva05}
{Ivanova}, N., {Belczynski}, K., {Fregeau}, J.~M., \& {Rasio}, F.~A. 2005,
  \mnras, 358, 572

\bibitem[{{Jayawardhana} {et~al.}(2006){Jayawardhana}, {Coffey}, {Scholz},
  {Brandeker}, \& {van Kerkwijk}}]{jay06}
{Jayawardhana}, R., {Coffey}, J., {Scholz}, A., {Brandeker}, A., \& {van
  Kerkwijk}, M.~H. 2006, \apj, submitted

\bibitem[{{Jilinski} {et~al.}(2005){Jilinski}, {Ortega}, \& {de la
  Reza}}]{jil05}
{Jilinski}, E., {Ortega}, V.~G., \& {de la Reza}, R. 2005, \apj, 619, 945

\bibitem[{{Kirkpatrick} {et~al.}(1999){Kirkpatrick}, {Reid}, {Liebert},
  {Cutri}, {Nelson}, {Beichman}, {Dahn}, {Monet}, {Gizis}, \&
  {Skrutskie}}]{kir99}
{Kirkpatrick}, J.~D., {Reid}, I.~N., {Liebert}, J., {Cutri}, R.~M., {Nelson},
  B., {Beichman}, C.~A., {Dahn}, C.~C., {Monet}, D.~G., {Gizis}, J.~E., \&
  {Skrutskie}, M.~F. 1999, \apj, 519, 802

\bibitem[{{K{\"o}hler} \& {Petr-Gotzens}(2002)}]{koh02}
{K{\"o}hler}, R., \& {Petr-Gotzens}, M.~G. 2002, \aj, 124, 2899

\bibitem[{{Lawson} {et~al.}(2001){Lawson}, {Crause}, {Mamajek}, \&
  {Feigelson}}]{law01}
{Lawson}, W.~A., {Crause}, L.~A., {Mamajek}, E.~E., \& {Feigelson}, E.~D. 2001,
  \mnras, 321, 57

\bibitem[{{Lawson} {et~al.}(2002){Lawson}, {Crause}, {Mamajek}, \&
  {Feigelson}}]{law02}
---. 2002, \mnras, 329, L29

\bibitem[{{Leinert} {et~al.}(1993){Leinert}, {Zinnecker}, {Weitzel},
  {Christou}, {Ridgway}, {Jameson}, {Haas}, \& {Lenzen}}]{lei93}
{Leinert}, C., {Zinnecker}, H., {Weitzel}, N., {Christou}, J., {Ridgway},
  S.~T., {Jameson}, R., {Haas}, M., \& {Lenzen}, R. 1993, \aap, 278, 129

\bibitem[{{Luhman}(2004)}]{luh04a}
{Luhman}, K.~L. 2004, \apj, 616, 1033

\bibitem[{{Luhman} {et~al.}(2003){Luhman}, {Stauffer}, {Muench}, {Rieke},
  {Lada}, {Bouvier}, \& {Lada}}]{luh03}
{Luhman}, K.~L., {Stauffer}, J.~R., {Muench}, A.~A., {Rieke}, G.~H., {Lada},
  E.~A., {Bouvier}, J., \& {Lada}, C.~J. 2003, \apj, 593, 1093

\bibitem[{{Luhman} \& {Steeghs}(2004)}]{luh04b}
{Luhman}, K.~L., \& {Steeghs}, D. 2004, \apj, 609, 917

\bibitem[{{Lyo} {et~al.}(2004){Lyo}, {Lawson}, {Feigelson}, \&
  {Crause}}]{lyo04b}
{Lyo}, A.-R., {Lawson}, W.~A., {Feigelson}, E.~D., \& {Crause}, L.~A. 2004,
  \mnras, 347, 246

\bibitem[{{Lyo} {et~al.}(2006){Lyo}, {Song}, {Lawson}, {Bessell}, \&
  {Zuckerman}}]{lyo06}
{Lyo}, A.-R., {Song}, I., {Lawson}, W.~A., {Bessell}, M.~S., \& {Zuckerman}, B.
  2006, \mnras, 392

\bibitem[{{Mamajek} {et~al.}(1999){Mamajek}, {Lawson}, \& {Feigelson}}]{mam99}
{Mamajek}, E.~E., {Lawson}, W.~A., \& {Feigelson}, E.~D. 1999, \apjl, 516, L77

\bibitem[{{Mamajek} {et~al.}(2000){Mamajek}, {Lawson}, \& {Feigelson}}]{mam00}
---. 2000, \apj, 544, 356

\bibitem[{{Mathieu} {et~al.}(2000){Mathieu}, {Ghez}, {Jensen}, \&
  {Simon}}]{mat00}
{Mathieu}, R.~D., {Ghez}, A.~M., {Jensen}, E.~L.~N., \& {Simon}, M. 2000,
  Protostars and Planets IV, 703

\bibitem[{{Meeus}(1991)}]{mee91}
{Meeus}, J. 1991, {Astronomical algorithms} (Willmann-Bell, Inc.)

\bibitem[{{Moffat}(1969)}]{mof69}
{Moffat}, A.~F.~J. 1969, \aap, 3, 455

\bibitem[{{Padgett} {et~al.}(1997){Padgett}, {Strom}, \& {Ghez}}]{pad97}
{Padgett}, D.~L., {Strom}, S.~E., \& {Ghez}, A. 1997, \apj, 477, 705

\bibitem[{{Patience} \& {Duch{\^e}ne}(2001)}]{pat01}
{Patience}, J., \& {Duch{\^e}ne}, G. 2001, in IAU Symposium, ed. H.~{Zinnecker}
  \& R.~{Mathieu}, 181--+

\bibitem[{{Peck} \& {Reeder}(1972)}]{pec72}
{Peck}, E.~R., \& {Reeder}, K. 1972, Journal of the Optical Society of America
  (1917-1983), 62, 958

\bibitem[{{Petr} {et~al.}(1998){Petr}, {Coude Du Foresto}, {Beckwith},
  {Richichi}, \& {McCaughrean}}]{pet98}
{Petr}, M.~G., {Coude Du Foresto}, V., {Beckwith}, S.~V.~W., {Richichi}, A., \&
  {McCaughrean}, M.~J. 1998, \apj, 500, 825

\bibitem[{{Reipurth}(2000)}]{rei00}
{Reipurth}, B. 2000, \aj, 120, 3177

\bibitem[{{Roe}(2002)}]{roe02}
{Roe}, H.~G. 2002, \pasp, 114, 450

\bibitem[{{Song} {et~al.}(2004){Song}, {Zuckerman}, \& {Bessell}}]{son04}
{Song}, I., {Zuckerman}, B., \& {Bessell}, M.~S. 2004, \apj, 600, 1016

\bibitem[{{Storer} \& {Kim}(1990)}]{sto90}
{Storer}, B.~E., \& {Kim}, C. 1990, Journal of the American Statistical
  Association, 85, 146

\end{thebibliography}

\end{document}